\documentclass[twocolumn]{aastex62}

\usepackage{amssymb}
\usepackage{amsmath}
\usepackage{latexsym}
\usepackage{natbib}
\usepackage{wasysym}
\usepackage{mathtools}
\usepackage[hang,flushmargin]{footmisc} 
\usepackage{colortbl}

\newcommand{\pb}{Kepler-80b} 
\newcommand{\pc}{Kepler-80c}
\newcommand{\pd}{Kepler-80d}
\newcommand{\pe}{Kepler-80e}
\newcommand{\pf}{Kepler-80f}
\newcommand{\pg}{Kepler-80g}
\newcommand{\kstar}{Kepler-80~}
\newcommand{\kstarn}{Kepler-80}

\definecolor{mgm}{rgb}{0.0, 0.0, 1.0}

\DeclareMathAlphabet{\mathpzc}{OT1}{pzc}{m}{it}

\begin{document}

\pagenumbering{arabic}

\shorttitle{Kepler-80 Resonant Chain}
\shortauthors{MacDonald et al.}

\correspondingauthor{Mariah G. MacDonald}
\email{mmacdonald@psu.edu}

\title{A Five-Planet Resonant Chain: Reevaluation of the Kepler-80 System}
\author[0000-0003-2372-1364]{Mariah G. MacDonald}
\affiliation{Department of Astronomy \& Astrophysics, Center for Exoplanets and Habitable Worlds, The Pennsylvania State University, University Park, PA 16802, USA}
\author[0000-0003-4709-2689]{Cody J. Shakespeare}
\affiliation{Department of Astronomy \& Astrophysics, Center for Exoplanets and Habitable Worlds, The Pennsylvania State University, University Park, PA 16802, USA}
\affiliation{Department of Physics and Astronomy, University of Nevada, Las Vegas, 4505 S. Maryland Pkwy, Las Vegas 89154, USA}
\author[0000-0003-1080-9770]{Darin Ragozzine}
\affiliation{Department of Physics and Astronomy, Brigham Young University, N283 ESC, Provo, UT 84602, USA}

\setcounter{footnote}{0}
\begin{abstract}
Since the launch of the Kepler space telescope in 2009 and the subsequent K2 mission, hundreds of multi-planet systems have been discovered. The study of such systems, both as individual systems and as a population, leads to a better understanding of planetary formation and evolution. Kepler-80, a K-dwarf hosting six super-Earths, was the first system known to have four planets in a chain of resonances, a repeated geometric configuration. Transiting planets in resonant chains can enable us to estimate not only the planets' orbits and sizes but also their masses. Since the original resonance analysis and TTV fitting of Kepler-80, a new planet has been discovered whose signal likely altered the measured masses of the other planets. Here, we determine masses and orbits for all six planets hosted by Kepler-80 by direct forward photodynamical modeling of the lightcurve of this system. We then explore the resonant behaviour of the system. We find that the four middle planets are in a resonant chain, but that the outermost planet only dynamically interacts in $\sim14$\% of our solutions. We also find that the system and its dynamic behaviour are consistent with \emph{in situ} formation and compare our results to two other resonant chain systems, Kepler-60 and TRAPPIST-1.
\end{abstract}




\section{Introduction}

A wealth of knowledge has already been obtained from the Kepler and K2 missions that has revolutionized the field of exoplanets. Subsequent analysis has discovered new classes of planets and systems grandly different from our own Solar System \citep[e.g., Kepler-11, ][]{Lissauer2011}. 

Although we are able to constrain planetary radii from the transit method employed by Kepler, much more work is required to recover mass information from the planets with meaningful constraints only possible in a fraction of systems. With estimates of both a planet's size and mass, we are able start exploring its formation history and its composition. A mass estimate requires radial velocity follow-up or for the planets to be gravitationally perturbing each other's orbits enough that we can detect significant variations in the time of transit, or TTVs. Many previous studies have examined TTVs ans successfully derived planets' masses \citep[e.g., ][]{hadden2017}. Fitting a system's TTVs can lead to determination of the planets' densities and eccentricities, making it one of the best ways to study a system.

One step beyond fitting the system's TTVs requires a self-consistent forward modeling of the system that fits directly to the lightcurve itself. Known as photodynamical modelling, it couples an n-body integrator with a limb-darkened transit model, skipping the requirement of measuring individual transit times \citep{Ragozzine2010}. This is particularly valuable in the case of low signal-to-noise transits like we have in Kepler-80. We use the PhotoDynamical Multiplanet Model -- PhoDyMM\url{https://github.com/dragozzine/PhoDyMM} -- developed by Ragozzine et al., in prep. and used on many past systems. For example, \citet{Mills2017} fit the lightcurve of Kepler-444 with an early version of PhoDyMM, constraining two planet masses and the orbital elements for all planets. 

\kstar is a K-dwarf hosting six super-Earth planets. The two largest planets were first discovered via TTVs by \citet{Xie2013} with orbital periods of 7.05 and 9.52 days. Two more planets were later validated by \citet{Lissauer2014} and \citet{Rowe2014}, and the innermost planet \pf~was statistically validated by \citet{Morton2016} with orbital periods of 3.07, 4.64, and 0.99 days, respectively. By fitting the TTVs of the four outer planets \footnote{The inner planet f is not dynamically interacting with the other planets and therefore does not exhibit TTVs}, \citet{MacDonald2016} constrained the orbital parameters and the masses of these four planets. They also studied the system's resonant behaviour, as \kstar was the first exoplanetary system with a confirmed four-body resonant chain.

More recently, \citet{Shallue2018} used neural nets and discovered a sixth planet with an orbital period of 14.3 days. Although a full analysis of the system's resonant behaviour was outside the scope of their study, they do recover a period ratio between planets c and g of near 3:2, suggesting that this new planet also participates in the resonant chain confirmed by \citet{MacDonald2016}. 

By directly forward modeling and fitting the lightcurve of this system with PhoDyMM, we aim to determine the masses and orbits for all six planets hosted by \kstar. In addition, we aim to investigate and characterize the resonant behaviour of the five outer planets and explore the formation of the system and its resonant chain.

In Section~\ref{sec:data}, we discuss our data and our methods. We describe the results of the PhoDyMM fitting in Section~\ref{sec:results} and characterize the system's resonant behaviour in Section~\ref{sec:resonances}. We verify that the system and its dynamic behaviour are consistent with \emph{in situ} formation in Section~\ref{sec:formation}. We then discuss our results and compare them to two other resonant systems, TRAPPIST-1 and Kepler-60, before summarizing and concluding in Section~\ref{sec:conclusion}.


\section{Data and Methods} \label{sec:data}

\subsection{Kepler Photometry}

We use all photometric data available by Kepler for this study, including 1 minute short cadence observations from Quarters 7, 8, 9, 11, 12, 13, 15, 16, and 17. Kepler-80 fell on Module 3 of the Kepler Space Telescope which suffered a failure early in the mission. Because of this, no data exist for Quarters 6, 7, and 14. The Kepler-80 lightcurve is detrended after masking the six known planets and stitched together following the methods described in Ragozzine et al. in prep. 

\subsection{Photodynamic Fitting}

To allow for the estimation of physical and orbital parameters of the small planets in Kepler-80 in a simultaneous and self-consistent manner, we fit the lightcurve directly instead of the system's TTVs. We use PhoDyMM which is described in detail in Ragozzine et al., in prep., which we summarize below. 

PhoDyMM integrates the Newtonian equations of motion for the star and the six planets. We then use a limb-darkened light-curve model to generate a synthetic lightcurve to compare to our data and measure a log likelihood assuming Gaussian uncertainies from Kepler lightcurve data. We then perform Bayesian parameter inference using Differential Evolution Markov Chain Monte Carlo \citep[DEMCMC, ][]{Ter2006}. For each planet, we fit the orbital period $P$, the mid-transit time $t_0$, the eccentricity $e$, the argument of periapse $\omega$, the sky-plane inclination $i$, the longitude of ascending node $\Omega$, the radius $R$, and the mass $M$. In addition, we fit for the star's mass $M_{\star}$ and radius $R_{\star}$, for the two limb-darkening coefficients $c_1$ and $c_2$, and for the amount of dilution from other nearby stars $d$.

We employ Gaussian priors on the stellar mass and radius based on values from \citet{MacDonald2016} ($M = 0.73\pm0.03 M_{\odot}$, $R=0.678\pm0.023 R_{\odot}$). We also fix $\Omega=0$ for all planets, given that the system seems to have small mutual inclinations. We employ flat priors on all other parameters, including a flat prior on the square root of the eccentricity. Nearly all parameters are well constrained by the data so we do not explore the effect of different priors. 



\section{Resulting Planetary Parameters}\label{sec:results}

We run a 96-chain DEMCMC with 190,000 steps, saving every 1,000 steps and removing a burn-in of 7,000 steps.  To assess long-term stability of our DEMCMC fits, we pull 30 random solutions from the DEMCMC posteriors and numerically integrate them for 100Myr using \texttt{REBOUND} \citep{rebound}. We find that all 30 solutions are long-term stable. We report the median values and 1$\sigma$ confidence intervals for the results from our photodynamic model in Table~\ref{tab:results}. 

\begin{deluxetable*}{lcccccc}
\tabletypesize{\footnotesize}
\tablecolumns{7}
\tablewidth{0pt}
\tablecaption{ Resulting Planetary Parameters \label{tab:results}}
\tablehead{
\colhead{Parameter} & \colhead{\pf} & \colhead{\pd} & \colhead{\pe} & \colhead{\pb} & \colhead{\pc} & \colhead{\pg} 
}
\startdata
$P$ (days) & $  0.98678^{+0.00001} _{-0.00002} $ & $  3.0723 ^{+0.0002} _{-0.0001} $ & $    4.6447^{+0.0001} _{-0.0002} $  & $    7.0534 \pm 0.0002 $  & $  9.5231\pm0.0001  $ &  $   14.651\pm0.001 $  \\
$R_p~(R_{\oplus})$  & $ 1.031 ^{+0.033}_{-0.027} $ & $1.309 ^{+0.036}_{ -0.032}$ & $1.330 ^{+0.039 }_{-0.038 }$  & $2.367 ^{+0.055 }_{-0.052}$  & $2.507 ^{+0.061}_{-0.058}$ &  $1.05 ^{+0.22}_{-0.24}$  \\
$M_p~(M_{\oplus})$ & 
  ---  & $5.95 ^{+0.65}_{-0.60} $ & $2.97 ^{+0.76 }_{-0.65} $  & $3.50 ^{+0.63}_{-0.57} $  & $3.49 ^{+0.63}_{-0.57} $ &  $0.065 ^{+0.044 }_{-0.038} $  \\
$\rho_p$ & --- & $14.6 ^{+1.9}_{-1.7} $ & $6.9 ^{+1.9 }_{-1.6} $  & $1.45 ^{+0.33}_{-0.29} $  & $1.22 ^{+0.24}_{-0.21} $ &  $0.31 ^{+0.46 }_{-0.20} $  \\
$e$ & $ 0.186^{+0.083 }_{-0.049} $ & $  0.0041^{+0.0037 }_{-0.0028} $ & $  0.0035^{+0.0032 }_{-0.0024} $  & $  0.0049^{+0.0036 }_{-0.0032} $  & $ 0.0079^{+0.0040}_{-0.0037} $ &  $ 0.1303^{+0.0034}_{-0.0037} $  \\
$i$ (deg)  &$85.99 ^{+0.48 }_{-0.52}$ & $89.24 ^{+0.46 }_{-0.37}$ & $88.59 ^{+0.15 }_{-0.16}$  & $88.989 ^{+0.090 }_{-0.085}$  & $88.744 ^{+0.049 }_{-0.046}$ &  $88.26 ^{+0.15 }_{-0.07}$  \\
$t_0$ (BJD - 2454900)  & $800.33893 ^{+0.00047 } _{-0.00055 }$ & $795.13111 ^{+0.00064 } _{-0.00067}$ & $796.9113 \pm 0.0011 $  & $758.39386 ^{+0.00097 } _{-0.0011 }$  & $796.04731 ^{+0.00049 } _{-0.00052} $ &  $758.5864 ^{+0.0094 } _{-0.0094}$  
\enddata
\tablecomments{Orbital period $P$, planetary radius $R_p$, planetary mass $M_p$, bulk density $\rho_p$, eccentricity $e$, sky-plane inclination $i$, and mid-transit time $t_0$ estimates resulting from the photodynamic DEMCMC posteriors. The nominal value for each parameter is the median of the posteriors and the lower and upper uncertainties are the 16th and 84th percentile confidence intervals. }
\end{deluxetable*}

\begin{figure}
    \centering
    \includegraphics[width=0.48\textwidth]{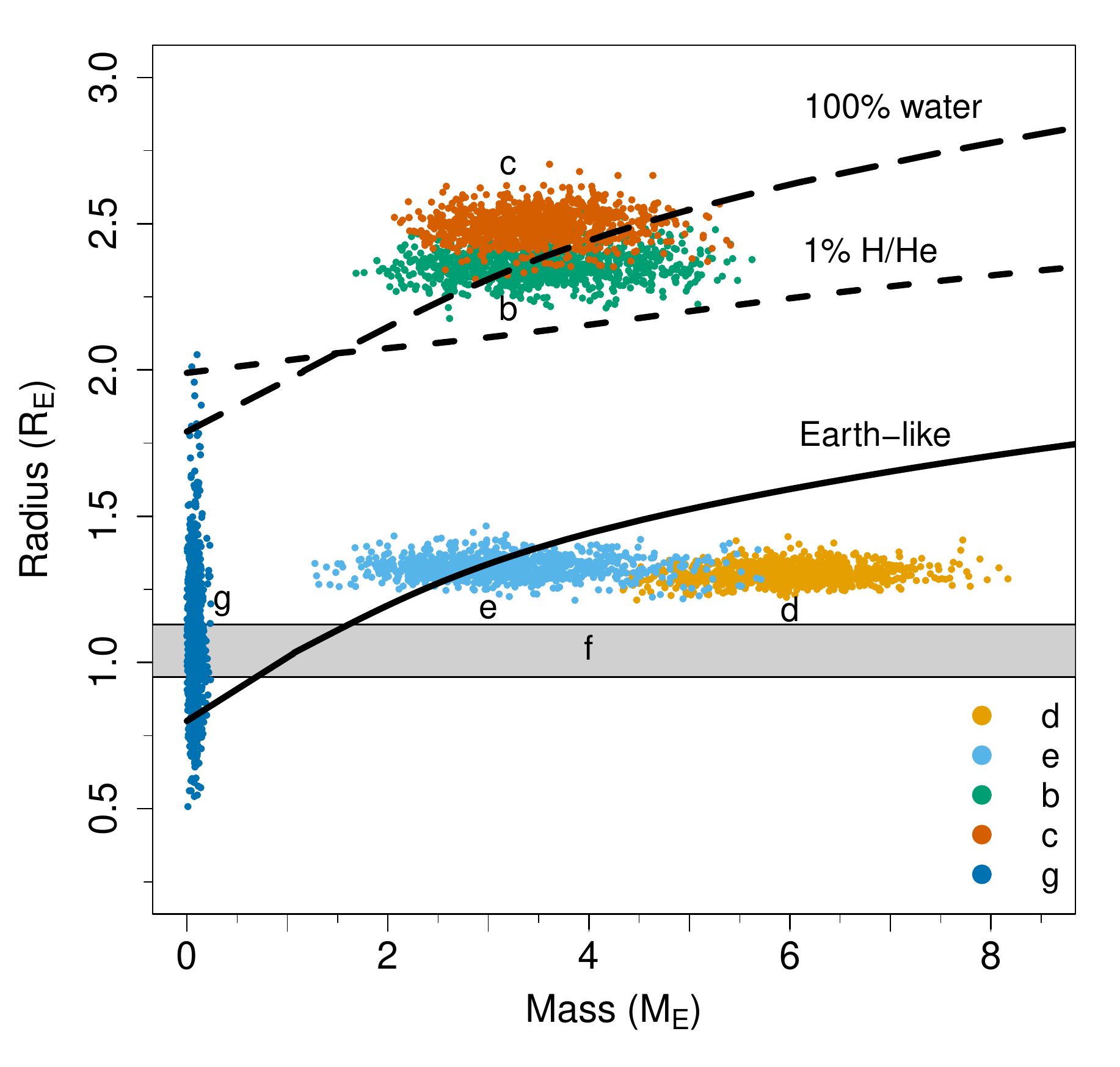}
    \caption{Mass-radius diagram with resulting estimates of Kepler-80's five outer planets. Instead of the typical points and error bars, we pull 200 points from our DEMCMC fitting. Since \pf~is not gravitationally interacting with the other planets, we are unable to constrain its mass and instead plot the 3$\sigma$ range of its radius. Our results are consistent with the measurements from \citet{MacDonald2016} to the 3$\sigma$ level. We estimate smaller masses for the planets, especially for \pb~and \pc, but estimate similar densities. }
    \label{fig:mr}
\end{figure}

Compared to the masses estimated by \citet{MacDonald2016} ($6.75^{+0.69}_{-0.51}$, $6.75^{+0.69}_{-0.51}$, $6.75^{+0.69}_{-0.51}$, $6.75^{+0.69}_{-0.51}~M_{\oplus}$), our new estimates for the four middle planets of $5.95 ^{+0.65}_{-0.60} $, $2.97 ^{+0.76 }_{-0.65} $,  $3.50 ^{+0.63}_{-0.57} $, and $3.49 ^{+0.63}_{-0.57} ~M_{\oplus}$ are more precise and smaller, suggesting that the methods we employed in \citet{MacDonald2016} were overestimating the masses. This overestimate, which is most prominent in the outer two planets \pb~and \pc, may be explained by the additional \pg~that was not modeled in \citet{MacDonald2016}. Since \pf~is not dynamically interacting with other planets, we are unable to constrain the planet's mass. Additionally, we estimate the mass of \pg~to be $0.065 ^{+0.044 }_{-0.038} ~M_{\oplus}$ and the radius to be $1.05 ^{+0.22}_{-0.24}~R_{\oplus}$. We show the mass and radius estimates for the five outer planets in Figure~\ref{fig:mr}. 

Given the degeneracy between the planets' masses and eccentricities, \citet{MacDonald2016} limited the eccentricities to less than 0.02. We measure nominal values of 0.0041, 0.0035, 0.0049, 0.0079, and 0.1303 for planets d, e, b, c, and g, respectively, which are consistent with the measurements from \citet{MacDonald2016} and confirm a dynamically cold system.

Our fits for Kepler-80g suggest that the planet has a fairly small mass ($\sim$0.6M$_{\mars}$) for its radius ($\sim$1.05R$_{\oplus}$), leading to a low bulk density. In addition, we find that the radius of this planet is poorly constrained while the mass appears to be well constrained. We postulate two reasons for this low density and varying precision. First, Kepler-80g has a comparatively low signal-to-noise ratio of 8.6 \citep{Shallue2018} compared to the much larger signatures of the other planets in the system \citep[40.4-92.3, ][]{MacDonald2016}, which can account for the poor constraint of the planet’s radius. Second, Kepler-80g is most likely not part of the resonant chain (see Section~\ref{sec:resonances}), leading to smaller gravitational perturbations with the neighboring planets. These smaller perturbations should lead to the planet’s mass being poorly constrained, but since it is fit with high precision, we conclude that we are most likely over-fitting this planet’s weak signal. Because of these two compounding issues, we caution the reader against drawing conclusions from our measured mass and density of Kepler-80g.

With both radius and mass estimates for our planets, we are able to calculate the planets' bulk densities and, from here, start to characterize the planets' compositions. We include composition curves for pure water, Earth-like, and 1\% H/He envelopes in our mass-radius diagram (see Figure~\ref{fig:mr}). Like \citet{MacDonald2016}, we find that \pd~and \pe~are consistent with an Earth-like composition, although \pd~ is likely more dense. Additionally, although we find smaller mass estimates for \pb~and \pc~than \citet{MacDonald2016}, we find consistent compositions that require substantial atmospheres of 1--2\% H/He. Given the poor constraint on \pg's radius, we are unable to constrain the planet's composition, although we note it is currently consistent with a terrestrial composition.

\begin{figure*}
    \centering
    \includegraphics[width=0.48\textwidth]{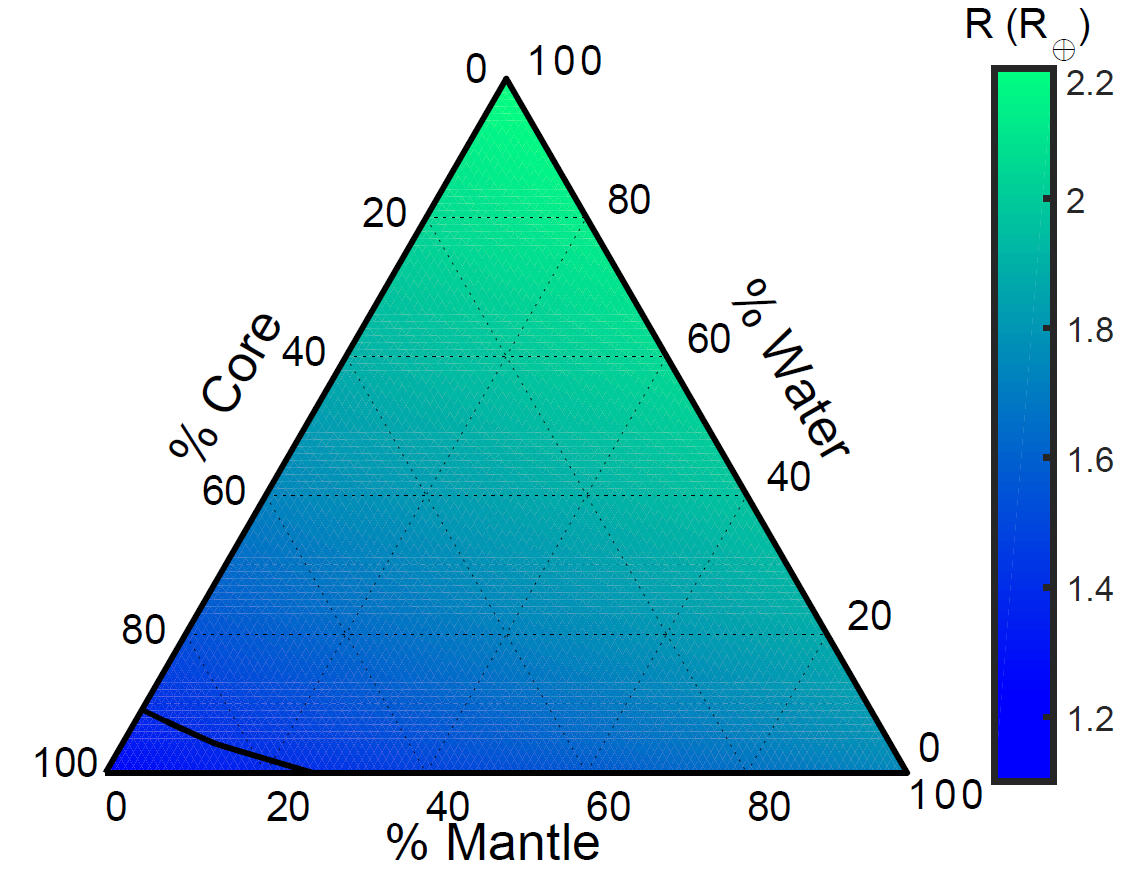}
    \includegraphics[width=0.48\textwidth]{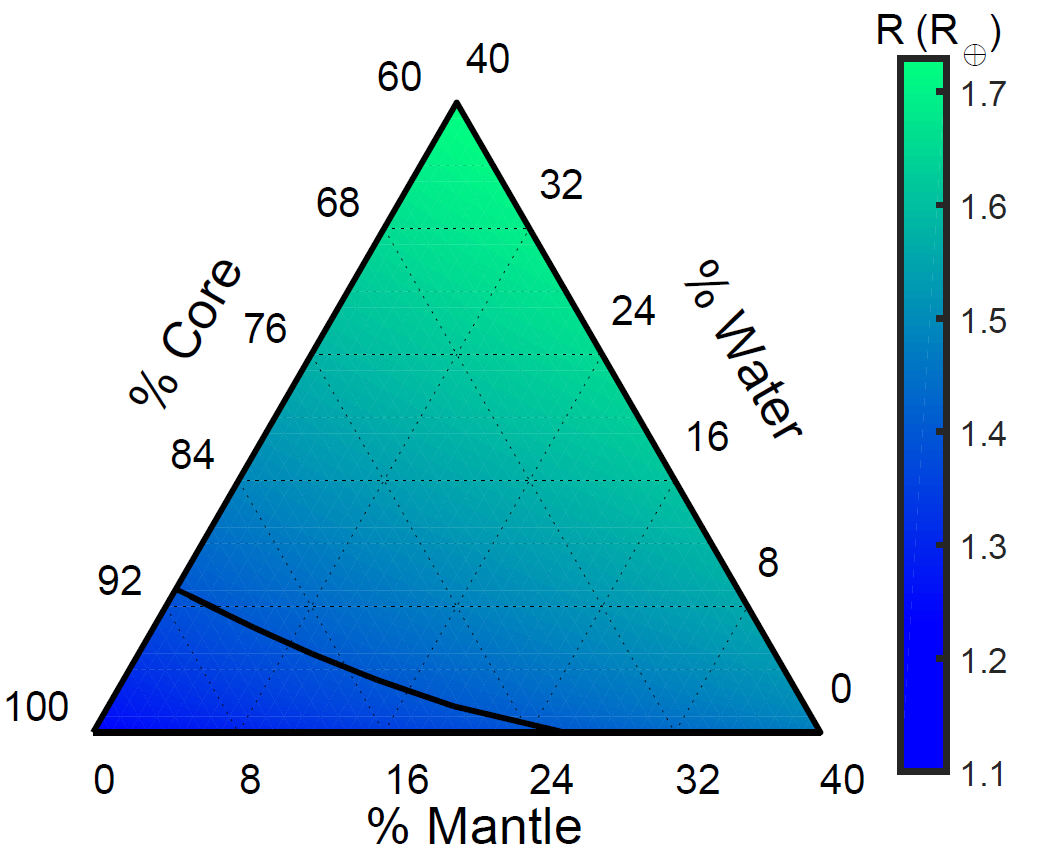}
    \caption{left) Ternary diagram from \pd~ MAGRATHEA models, where the axes are percentage of mass in a core, mantle, and water ocean. Here, the planet was fixed to a mass of 5.95$M_{\oplus}$, and the core and mantle are modeled to be pure liquid/solid iron and perovskite and post-perovskite, respectively. We colour the diagram based on the resulting planetary radius, ranging from a minimum radius of 1.15$R_{\oplus}$ to a maximum radius of 2.2$R_{\oplus}$. We include a contour line for R=1.44$R_{\oplus}$, the 3$\sigma$ upper limit of the radius. right) Ternary diagram of the same models, but zoomed in to core mass percentages of 60\% -- 100\%, and therefore with a change in the colourbar scale. We find that \pd~ must be at least 92\% core to satisfy its density.}
    \label{fig:comp_d}
\end{figure*}

\begin{figure*}
    \centering
    \includegraphics[width=0.48\textwidth]{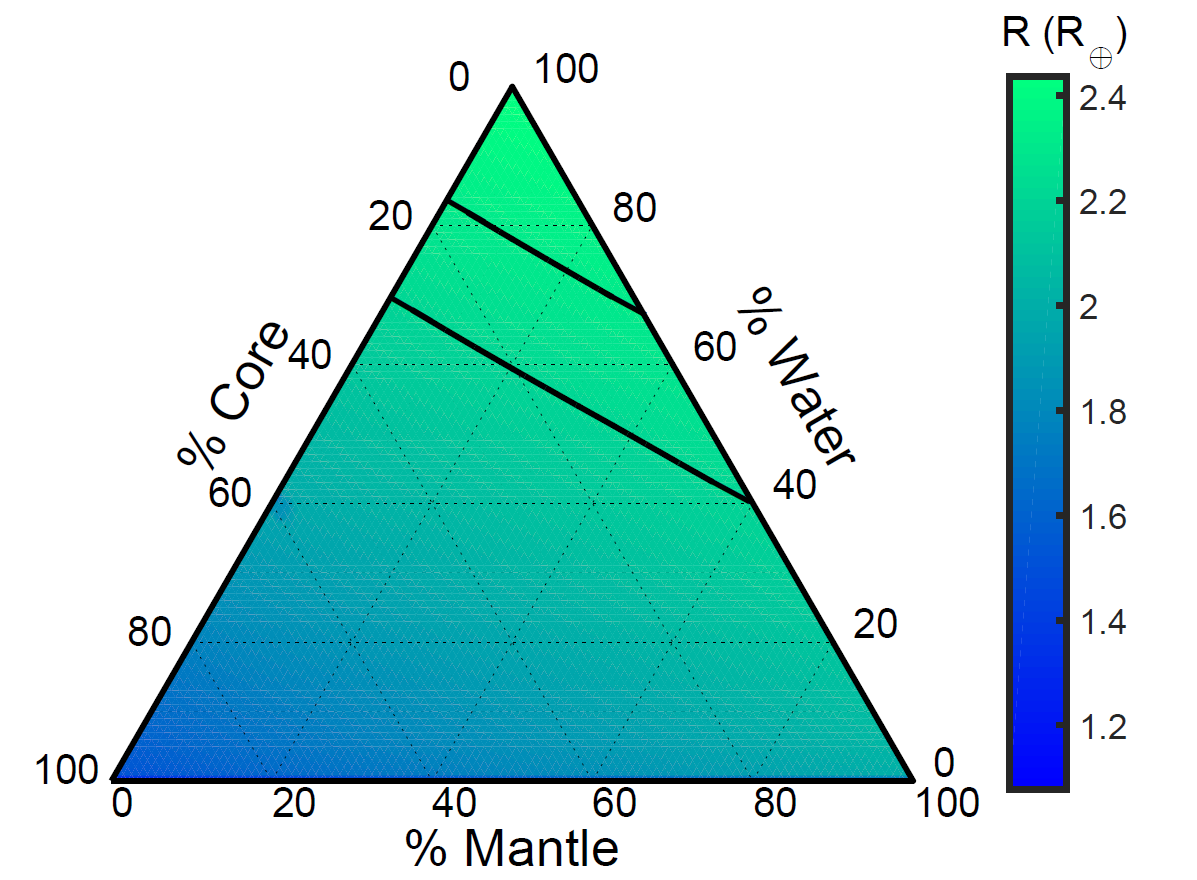}
    \caption{Ternary diagram from \pb~ MAGRATHEA models, where the axes are percentage of mass in a core, mantle, and water ocean, and we force 0.03\% mass into the atmosphere. Here, the planet was fixed to a mass of 3.5$M_{\oplus}$, and the core and mantle are modeled to be pure liquid/solid iron and perovskite and post-perovskite, respectively. We colour the diagram based on the resulting planetary radius, ranging from a minimum radius of 1.6$R_{\oplus}$ to a maximum radius of 2.42$R_{\oplus}$. We include contour lines at R=2.211$R_{\oplus}$ and R=2.315$R_{\oplus}$, the 1$\sigma$ and 3$\sigma$ lower limits of the radius. We find that \pb~ must contain at least 0.01\% mass it its atmosphere to satisfy its low density.}
    \label{fig:comp_b}
\end{figure*}

Kepler-80d's density is inferred to be quite high. We model the interiors of \pd~ and \pb~ using the planet structure code MAGRATHEA (Huang, Rice, and Steffen, in prep.). Here, the core is made of solid and liquid iron, and the mantle is made of perovskite and post-perovskite. To fully explore potential compositions, we model thousands of planets at their nominal masses ($M_d=5.95M_{\oplus}$, $M_b=3.5M_{\oplus}$) with integer percentages of mass in the planet's core, mantle, and a water ocean on top of the mantle. For \pb, we separate our models into suites that include an atmosphere with a mass of 0.0\%, 0.01\%,  and 0.03\%  of the planet's mass\footnote{For reference, Venus has an atmospheric mass fraction of $\sim0.01\%$}. We plot the resulting modeled radii as ternary plots in Figure~\ref{fig:comp_d} for \pd~ and Figure~\ref{fig:comp_b} for \pb. We find that while \pd~ must be at least 92\% core to satisfy its density, \pb~ requires a Venus-like atmosphere ($\sim$0.01\% mass).

\kstar is relatively faint (Kepler magnitude of 14.8), but may be of interest for future observational studies. We draw 100 samples from the posterior distribution and calculate transit times and durations. These are reported in Table \ref{tab:future}.

\begin{deluxetable*}{lllllll}
\tabletypesize{\footnotesize}
\tablecolumns{7}
\tablewidth{0pt}
\tablecaption{ Future Times and Durations \label{tab:future}}
\tablehead{
\colhead{Planet} & \colhead{Time} & \colhead{$t^+$} & \colhead{$t_-$} & \colhead{Duration} & \colhead{$D^+$} & \colhead{$D_-$} \\
\colhead{} & \colhead{(BJD-2454900)} & \colhead{min} & \colhead{min} & \colhead{hr} & \colhead{hr} & \colhead{hr}
}
\startdata
b	&	4503.90	&	58.39	&	-48.99	&	2.429	&	0.121	&	-0.109	\\
b	&	4510.96	&	58.44	&	-49.20	&	2.429	&	0.121	&	-0.109	\\
b	&	4518.01	&	59.14	&	-49.38	&	2.432	&	0.125	&	-0.111	\\
b	&	4525.07	&	59.28	&	-49.27	&	2.432	&	0.125	&	-0.111	\\
b	&	4532.12	&	59.53	&	-49.12	&	2.430	&	0.123	&	-0.109	\\
b	&	4539.18	&	59.86	&	-49.12	&	2.429	&	0.121	&	-0.110	\\
b	&	4546.23	&	61.46	&	-48.19	&	2.429	&	0.120	&	-0.111	\\
b	&	4553.28	&	61.90	&	-48.30	&	2.429	&	0.121	&	-0.110	\\
b	&	4560.34	&	62.51	&	-48.65	&	2.429	&	0.121	&	-0.111	\\
b	&	4567.39	&	62.76	&	-49.65	&	2.430	&	0.125	&	-0.113	\\
b	&	4574.45	&	63.66	&	-49.14	&	2.427	&	0.126	&	-0.112	\\
b	&	4581.50	&	63.94	&	-49.72	&	2.426	&	0.122	&	-0.111	\\
b	&	4588.55	&	63.91	&	-50.12	&	2.425	&	0.120	&	-0.111	\\
b	&	4595.60	&	63.84	&	-50.77	&	2.425	&	0.120	&	-0.111	\\
b	&	4602.65	&	63.97	&	-51.11	&	2.423	&	0.120	&	-0.111
\enddata
\tablecomments{Projected future times and durations of transits of the planets in the Kepler-80 system. Times and durations are the 50th percentile, uncertainties are estimated using the 16th and 84th percentiles, mimicking 1-$\sigma$ uncertainties. Only a portion of this table is shown here to demonstrate its form and content. A machine-readable version of the full table is available.}
\end{deluxetable*}


\section{Five-planet resonant chain}\label{sec:resonances}


\kstar earned the title of first exoplanetary system with a confirmed three-body resonant chain of longer than three planets. Before the discovery of the outermost planet g, \citet{MacDonald2016} confirmed and characterized the resonance between the four inner planets d, e, b, and c. These planets are locked in a resonant chain of 4:6:9:12 with two-body resonances between adjacent pairs of 3:2 (e:d), 3:2 (b:e), and 4:3 (c:b). \citet{Shallue2018} suggest that planet g is included in this chain, bringing the chain to 4:6:9:12:18 with a 3:2 resonance between planets \pg~and \pc. 

These two-body resonances are characterized by the libration of the two-body angles:

\begin{equation}
    \Theta_{1-2} = j_1\lambda_1 + j_2\lambda_2 + j_3\omega_1 + j_4\omega_2 + j_5\Omega_1+j_6\Omega_2,
\end{equation}
\noindent where 1 and 2 refer to two planets, $\lambda$ is the mean longitude, $\omega$ is the argument of periapse, $\Omega$ is the longitude of ascending node, and the $j$ coefficients must sum to zero.

In addition to being a set of consecutive two-body angles, a resonant chain can also be characterized by three-body angles, defined as:

\begin{equation}
    \phi = p\lambda_1 - (p+q)\lambda_2 + q\lambda_3
\end{equation}

\noindent where $\lambda_i$ is the planet's longitude, planet 1 is the inner planet in the trio, and $p$ and $q$ are coefficients describing the resonance.

\begin{figure*}
    \centering
    \includegraphics[width=0.48\textwidth]{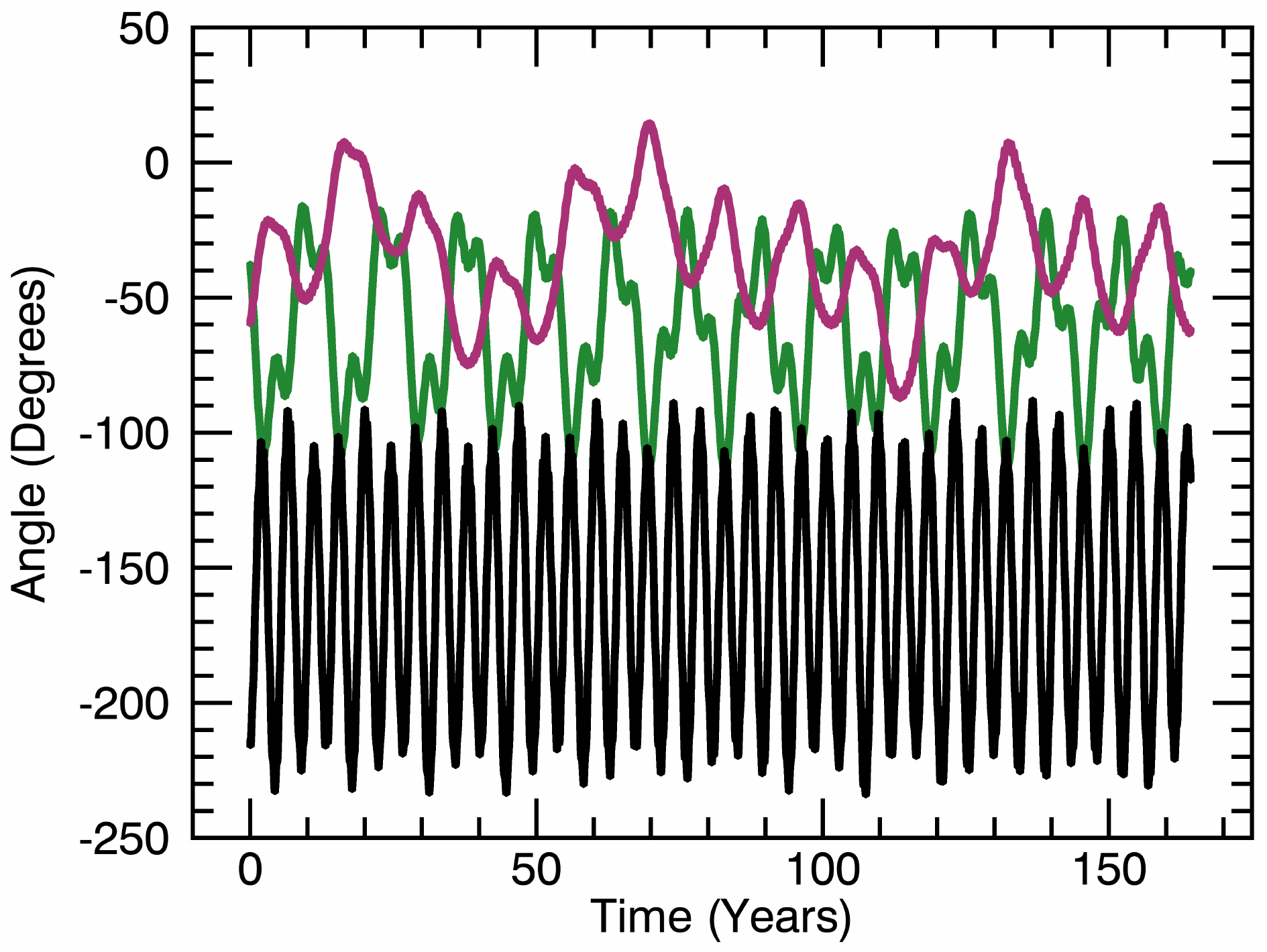}
    \includegraphics[width=0.48\textwidth]{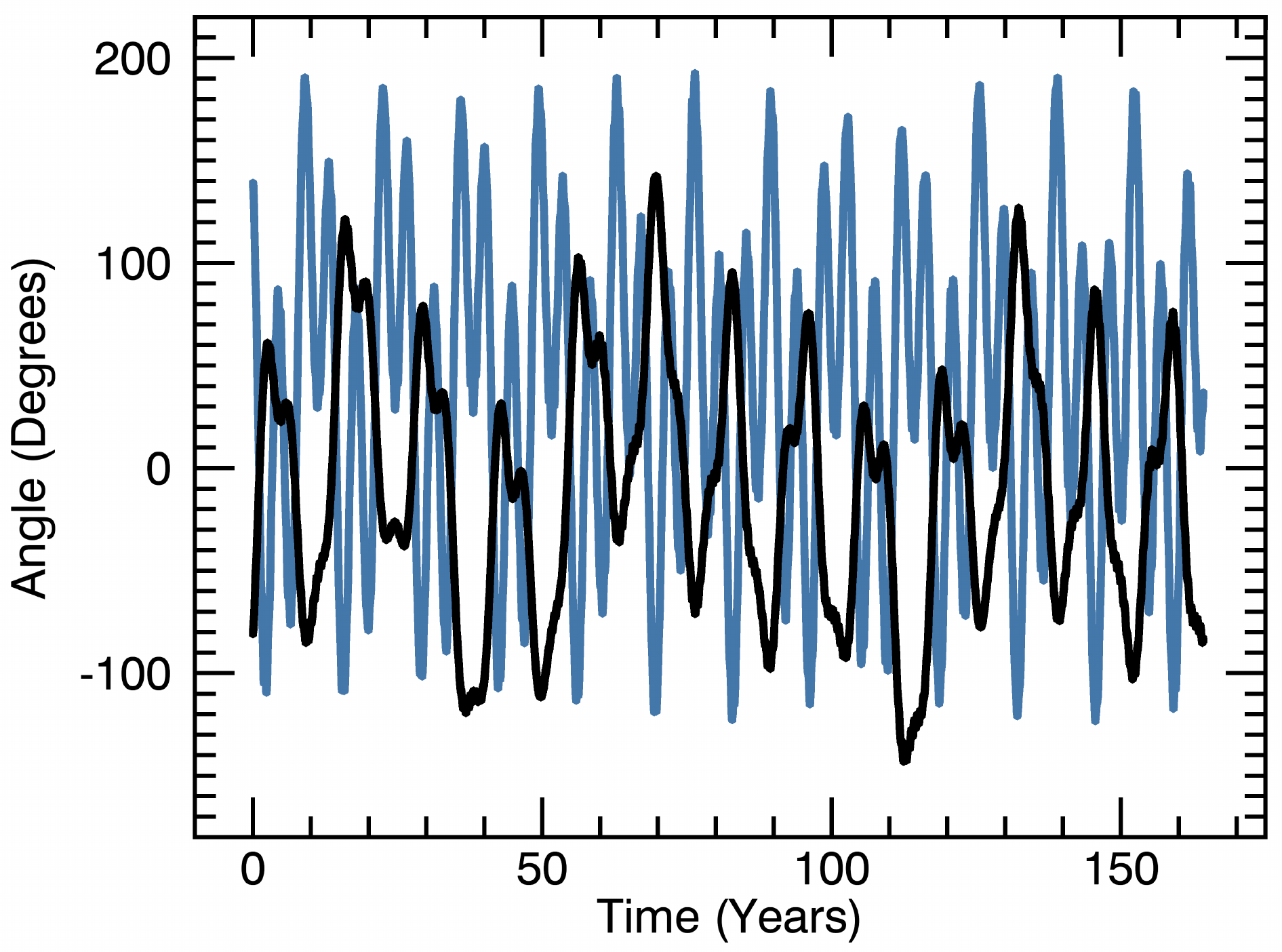}
    \caption{Left) Three-body and right) four-body angles from one of the photodynamic fits of Kepler-80. The three-body angles $\phi_1$, $\phi_2$, and $\phi_3$ are colored black, green, and purple, respectively. We plot the inner four-body angle $\phi_2-\phi_1$ in blue and the outer four-body angle  $\phi_3-\phi_2$ as black. }
    \label{fig:angles}
\end{figure*}

We pull 1000 fits from our photodynamic model and integrate the systems forward in time to study \kstarn's resonant behaviour. We find all three three-body angles, both four-body angles, and the five-body angle to be librating in some of these integrations. We show examples of these librations from one of the fits in Figure~\ref{fig:angles}. In addition, we find that the two-body angles from all adjacent planet pairs rarely librate. We summarize the centers and amplitudes of each of the two- and three-body angles, and the frequency of their libration, in Table~\ref{tab:res_angles}.

\begin{deluxetable}{lcll}
\tabletypesize{\footnotesize}
\tablecolumns{4}
\tablewidth{0pt}
\tablecaption{ Resonant Angles \label{tab:res_angles}}
\tablehead{
\colhead{Angle} & \colhead{\% libration} & \colhead{Center} & \colhead{Amplitude} 
}
\startdata
$\phi_1~=~3\lambda_b - 5\lambda_e + 2 \lambda_d$ & 100 & $200.6\pm3.1$ & $39.0\pm3.8$ \\
  &   & $180.0\pm3.9$ & $34.0\pm5.5$ \\
   &   & $161.9\pm4.4$ & $41.1\pm4.8$ \\
$\phi_2~=~2\lambda_c - 3\lambda_b + 1 \lambda_e$ & 100 & $-60.5\pm6.3$ & $16.4\pm5.5$ \\
  &  & $0.3\pm24.3$ & $56.1\pm13.7$ \\
    &   & $55.0\pm9.9$ & $24.3\pm5.0$ \\
$\phi_3~=~1\lambda_g - 2\lambda_c + 1 \lambda_b$ & 8.2 & $-33.8\pm42.3$ & $26.7\pm21.2$ \\
  & 46.3$^\dagger$ & $166.0\pm26.3$ & $20.5\pm15.7$ \\
$\Theta_{d,e}~=~3\lambda_e - 2\lambda_d - \bar{\omega_d}$ & 11.4 & $179.9\pm0.5$ & $32.6\pm6.5$ \\
$\Theta_{e,b}~=~3\lambda_b - 2\lambda_e - \bar{\omega_e}$ & 0.1 &   --- *  &   --- *  \\
$\Theta_{b,c}~=~3\lambda_c - 2\lambda_b - \bar{\omega_b}$ & 0 &   ---  &   ---  \\
$\Theta_{c,g}~=~3\lambda_g - 2\lambda_c - \bar{\omega_c}$ & 0 &   ---  &   ---  \\
\enddata
\tablecomments{Percentage of librating angles in 1000 randomly sampled DEMCMC bestfits with the associated centers and amplitudes for each resonant angle. We find that the initial chain reported by \citet{MacDonald2016} is always librating. We also find that all five dynamically interacting planets in this system are in a full resonant chain, with every planet in a librating two-, three-, four- or five-body angle, in 14\% of our bestfits. \newline * This angle liberated in one bestfit with a center of 163.5 and an amplitude of 34.4. }
\end{deluxetable}

The three-body angles librate about various centers, and the centers of $\phi_1$ and $\phi_2$ are strongly correlated. Our bestfit solutions favour libration centers of $\phi_1\sim200^{\circ}$, $\phi_2\sim-60^{\circ}$ (62.8\%), while 30.9\% result in $\phi_1\sim180.0^{\circ}$, $\phi_2\sim0^{\circ}$ and the remaining 6.3\% of bestfits have angles which librate about $\phi_1\sim162^{\circ}$, $\phi_2\sim55^{\circ}$. The primary libration center of $\phi_1\sim200^{\circ}$ agrees well with both observations \citep{MacDonald2016} and formation simulations \citep{MacDonald2018}. However, the libration center of $\phi_2\sim-60^{\circ}$ does not agree with previous studies ($\sim-72^{\circ}$). We find no correlations between libration centers and any planetary parameters, but this change in libration center could be due to \pg, which was unknown at the time of \citet{MacDonald2018}.

We find that planets d, e, b, and c are always in a resonant chain with the associated three-body angles librating in all of our bestfits. However, \pg~does not always participate in the chain. The three-body angle between planets b, c, and g and the four-body angle between planets e, b, c, and g only librate 8.2\% and 14.1\% of the time, respectively, and the two-body angle between planets c and g does not librate in any of the bestfits. However, since a resonant chain is defined by all of the planets interacting in the chain with a librating angle, even if not all of the angles librate, Kepler-80 very well could have a five-planet resonant chain.

To explore why the resonant angles are or are not librating, we perform the following analysis. We first look to see if the distributions of planet mass, eccentricity, and argument of periapse when the outer three-body angle is librating are statistically different than when the angle is not librating using Kolmogorov–Smirnov and Anderson-Darling two-sample tests. These tests result in large p-values for every planet mass and eccentricity, and so we fail to reject the null hypothesis that the two samples are from the same population.

From both Kolmogorov–Smirnov and Anderson-Darling two-sample tests, we find that \pe~ and \pc~have statistically distinct distributions of their arguments of periapse when $\phi_3$ is librating and when it is not librating. This suggests that there is a preferred orientation of the outer planet, most likely because it is eccentric ($e\sim0.13$).

\begin{figure*}
    \centering
    \includegraphics[width=0.37\textwidth]{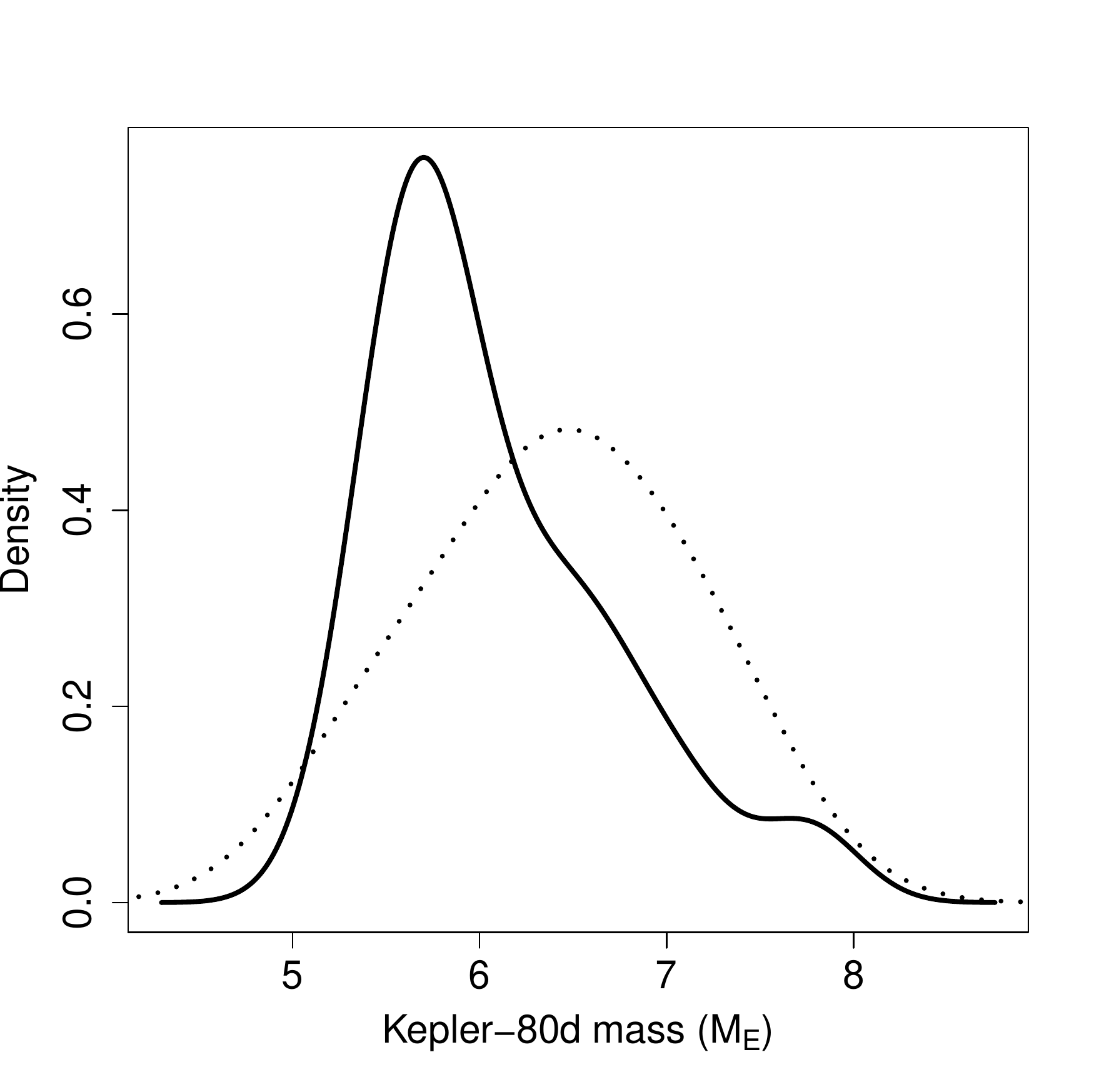}
    \includegraphics[width=0.37\textwidth]{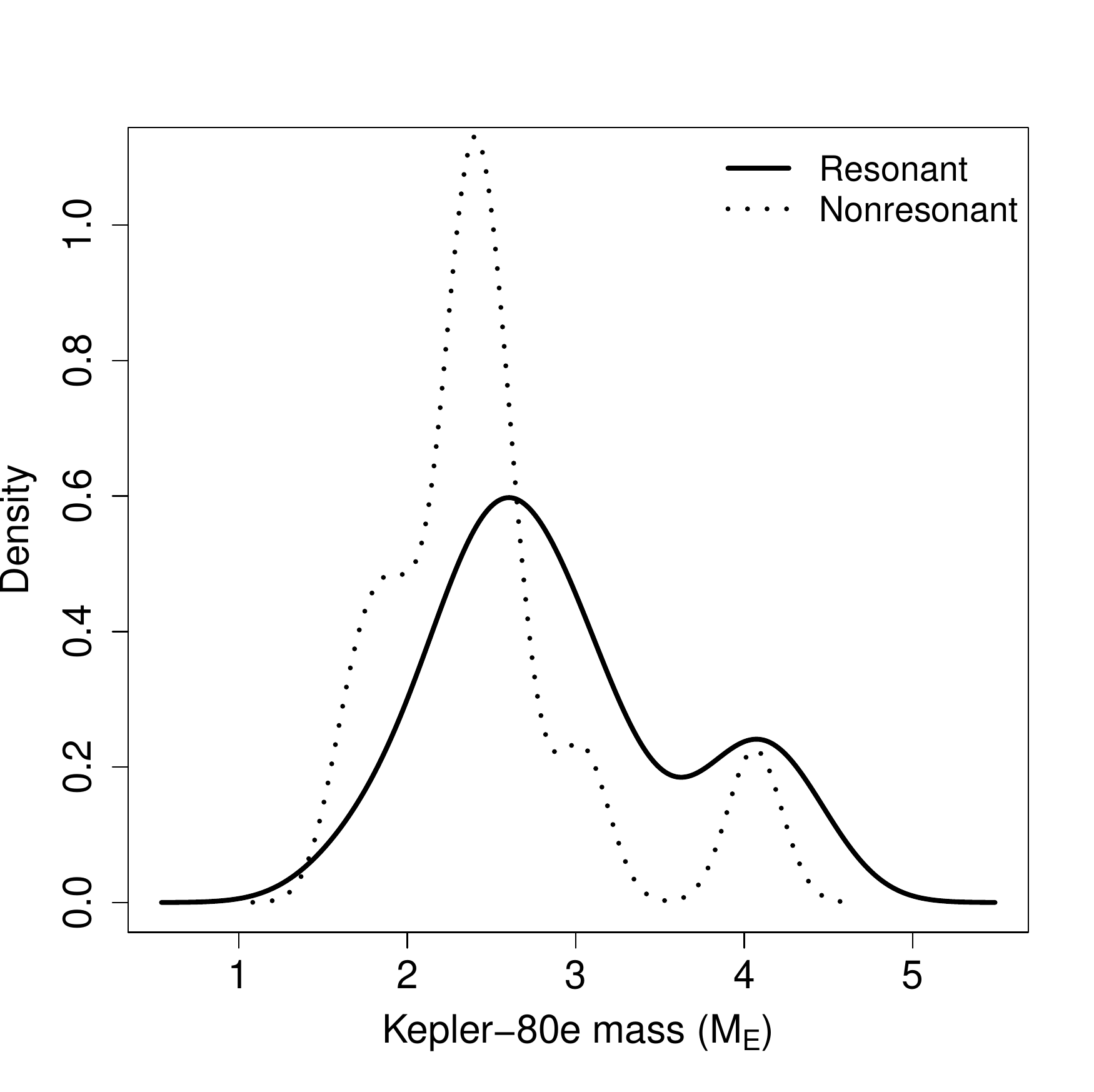}
    \includegraphics[width=0.37\textwidth]{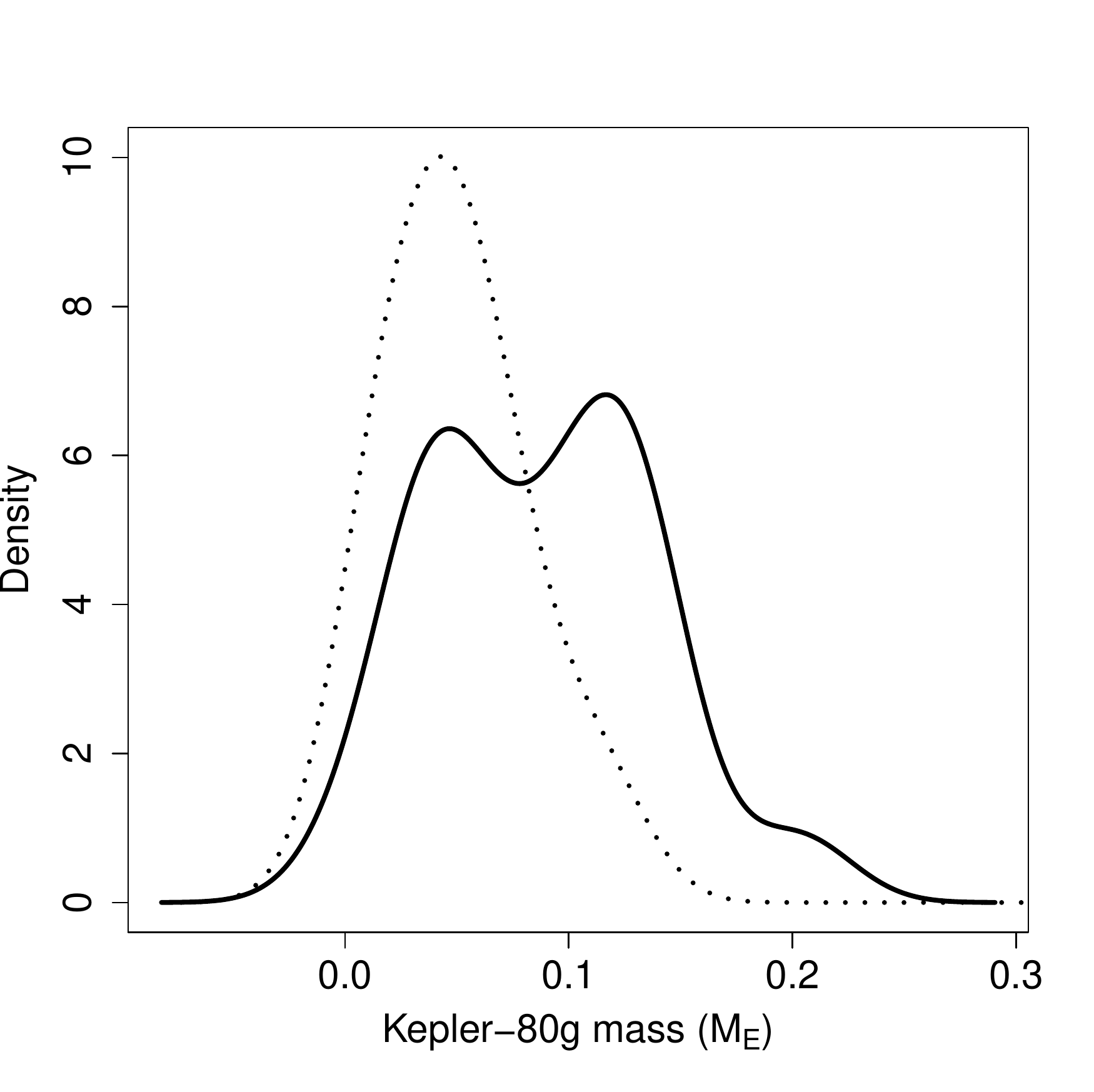}
    \includegraphics[width=0.37\textwidth]{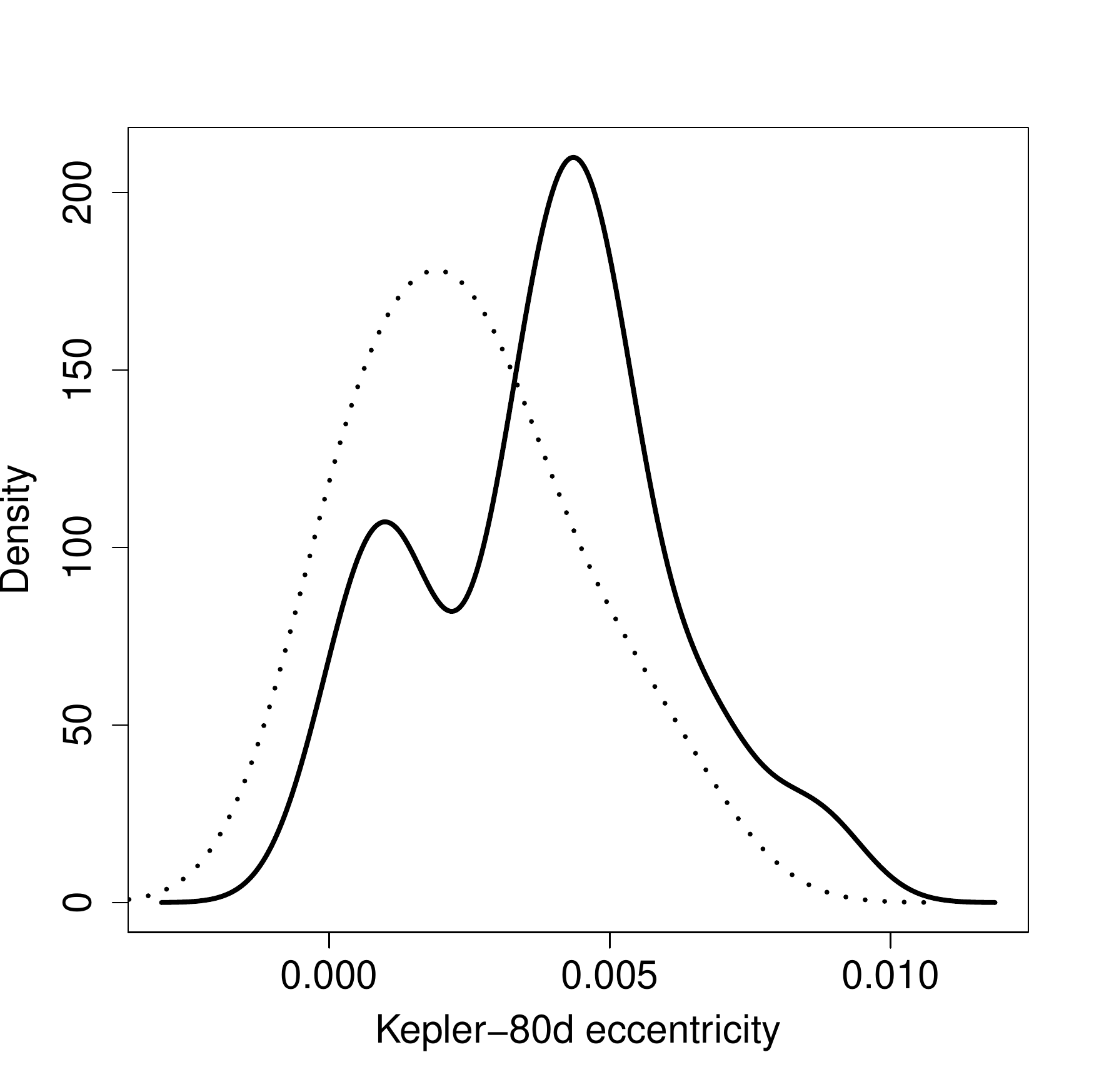}
    \includegraphics[width=0.37\textwidth]{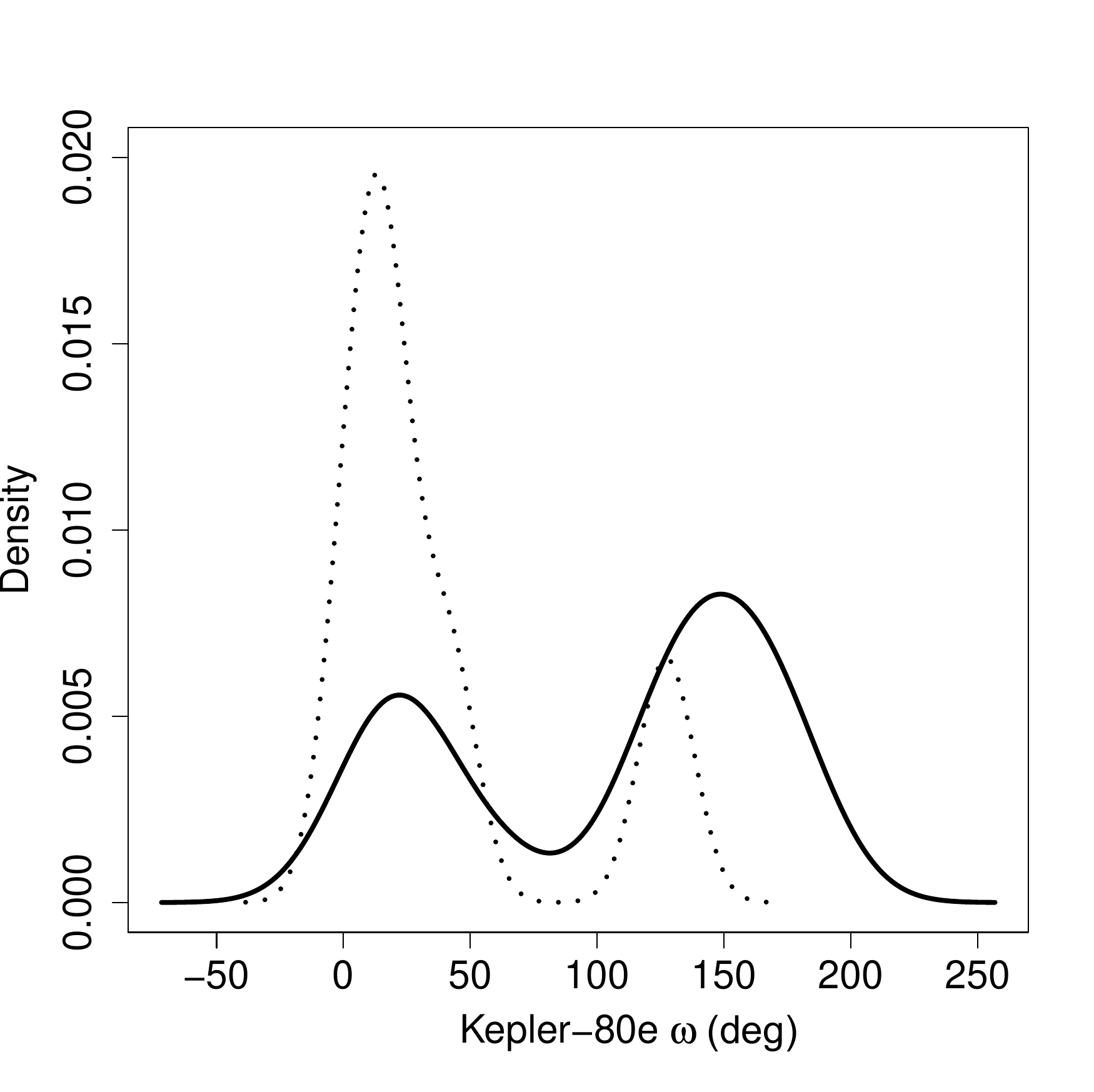}
    \includegraphics[width=0.37\textwidth]{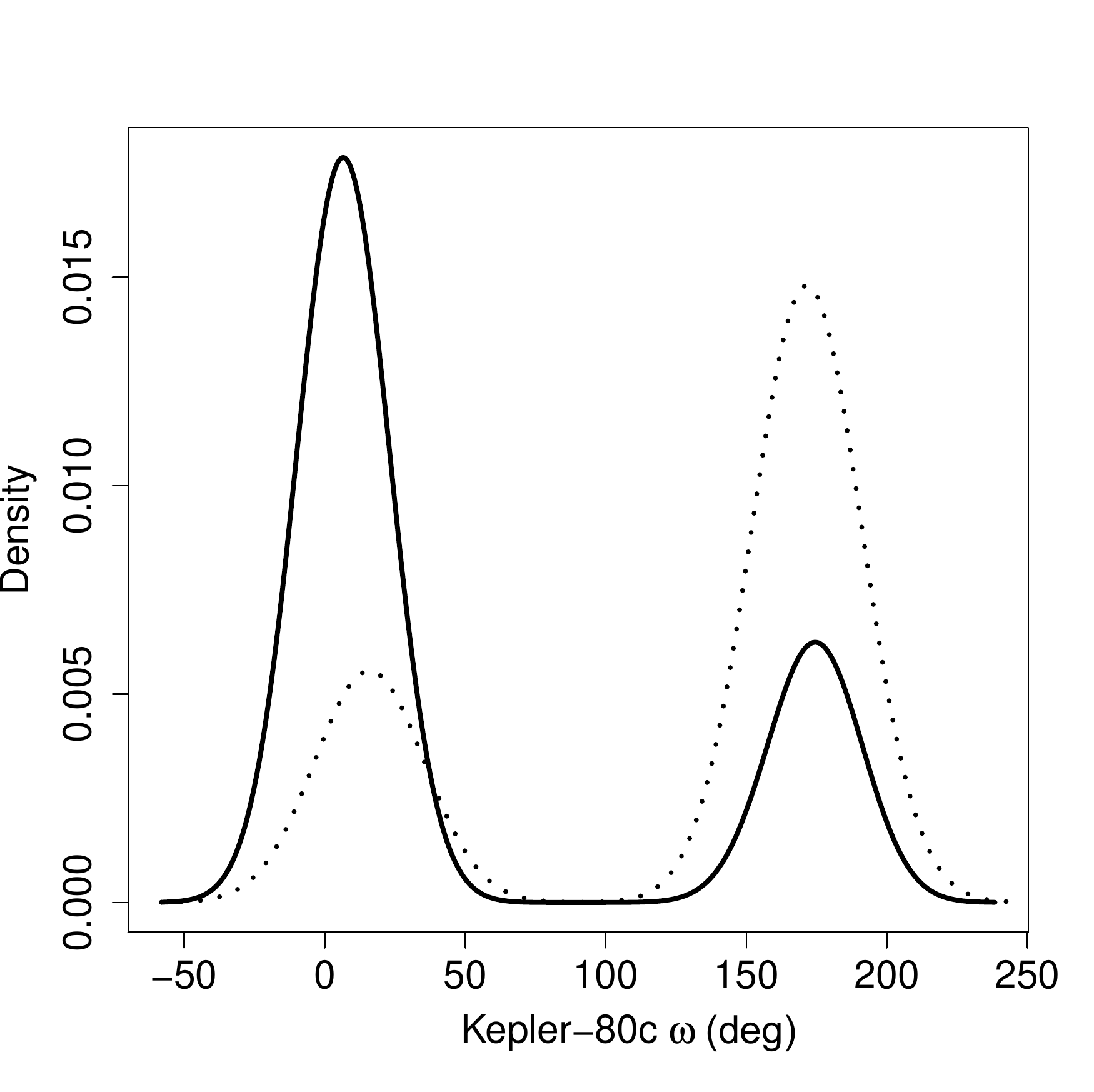}
    \caption{ Kernel Density Estimates of some mass, eccentricity, and argument of periapse ($\omega$) distributions for each planet, split by whether or not the outer three-body angle $\phi_3$ was librating. We find that, within our simulations, there is a slight preference for a less massive \pd, a more massive \pe, and a more massive \pg, however, we caution against drawing any conclusions from this, as the mass distributions are not statistically distinct. We also find that there is a slight preference for a more eccentric \pd, but again, these two distributions are not statistically distinct. Finally, we find there is a preference for the arguments of periapse of \pe~ and \pc~ to be $\sim150^{\circ}$ and $\sim0^{\circ}$, respectively. The distributions of the angles for these two planets are distinct, with Anderson-Darling two-sample test p-values of 0.007 and 0.005.}
    \label{fig:kde_whyres}
\end{figure*}

We then explore Kernel Density Estimates (KDE) of the mass, eccentricity, and argument of periapse distributions for each planet. We show some of these KDEs in Figure~\ref{fig:kde_whyres}. We find that, within our bestfits, there is a slight preference for a less massive \pd, a more massive \pe, and a more massive \pg~ for the outer three-body angle $\phi_3$ to librate. However, we caution against drawing any conclusions from this, as the mass distributions are not statistically distinct. We also find that there is a slight preference for a more eccentric \pd, but again, these two distributions are not statistically distinct. Finally, we find there is a preference for the arguments of periapse of \pe~ and \pc~ to be $\sim150^{\circ}$ and $\sim0^{\circ}$, respectively, and the distributions of the angles for these two planets are distinct.

\section{Kepler-80's Resonant Chain is Consistent with in Situ Formation}\label{sec:formation}

Following \citet{MacDonald2018}, we explore the formation of the resonant chain of \kstar via in situ formation using $N$-body simulations in \texttt{REBOUND} \citep{rebound}. A resonant chain can form in situ via two pathways: with small changes to the planets' semi-major axes (which we will call short-scale migration) and with small changes to the planets' eccentricities (which we will call eccentricity damping). We note that short-scale migration becomes pure eccentricity damping when the semi-major axis damping timescale $\tau_a$ is large.

\begin{figure*}
    \centering
    \includegraphics[width=0.48\textwidth]{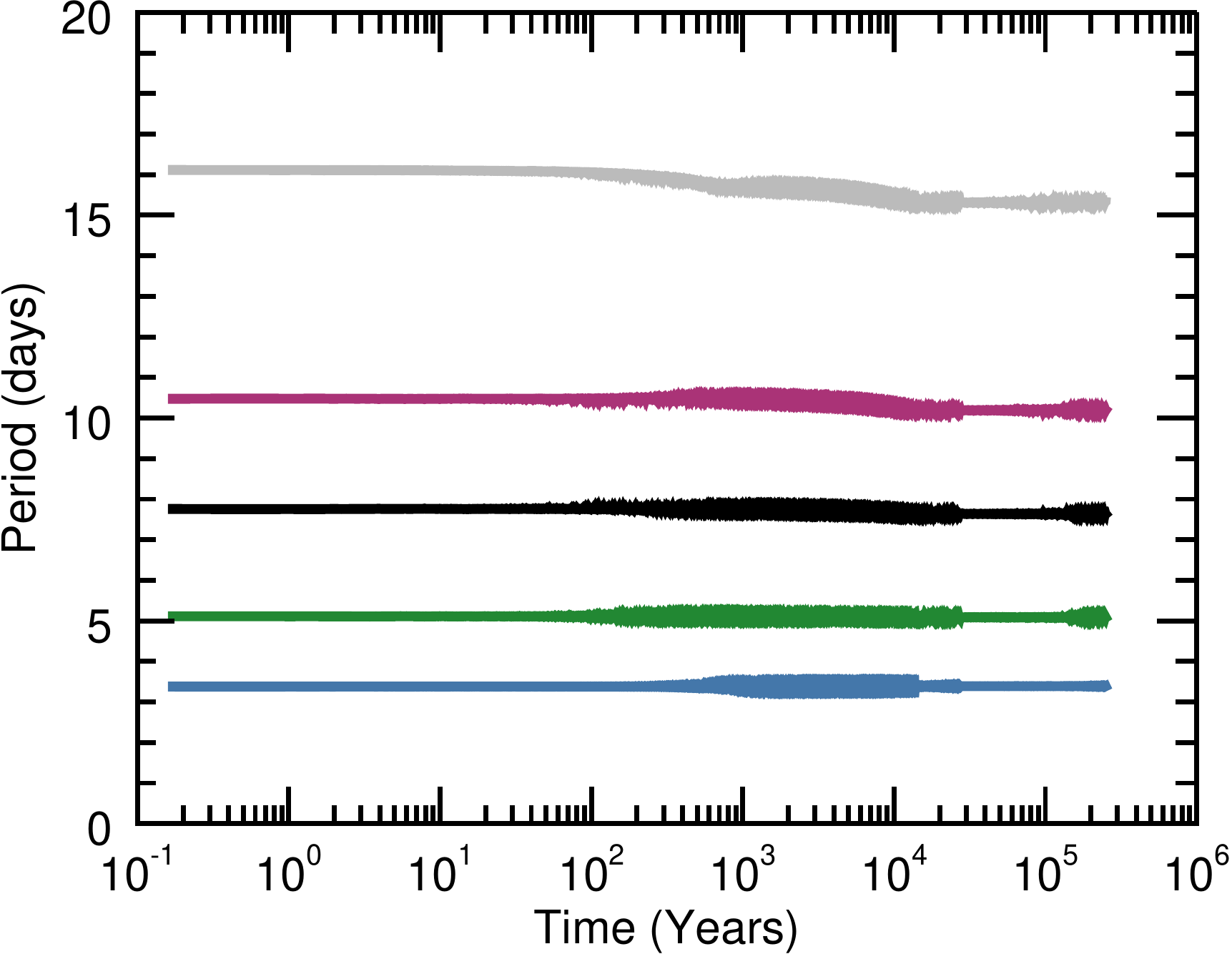}
    \includegraphics[width=0.48\textwidth]{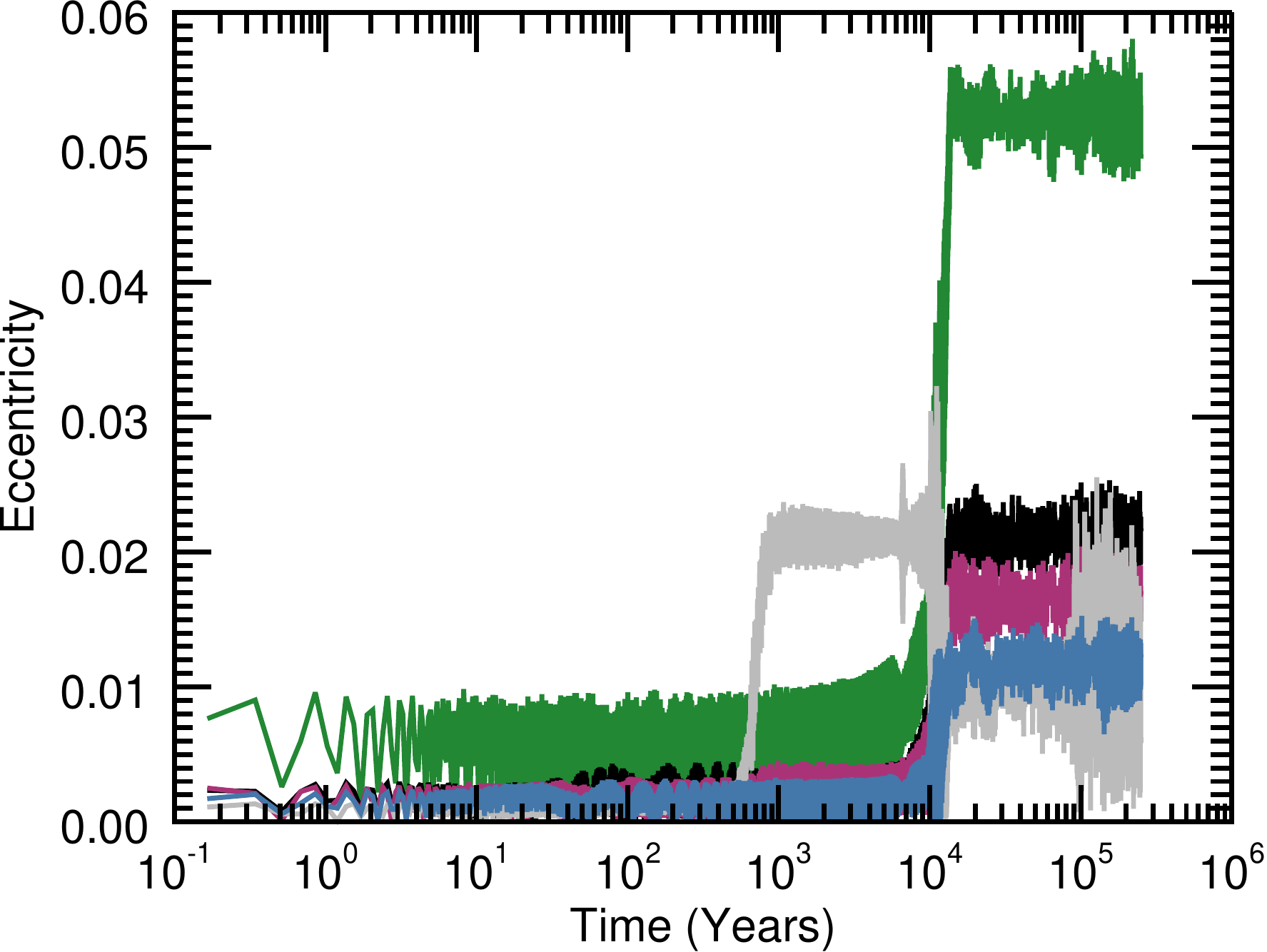}
    \includegraphics[width=0.48\textwidth]{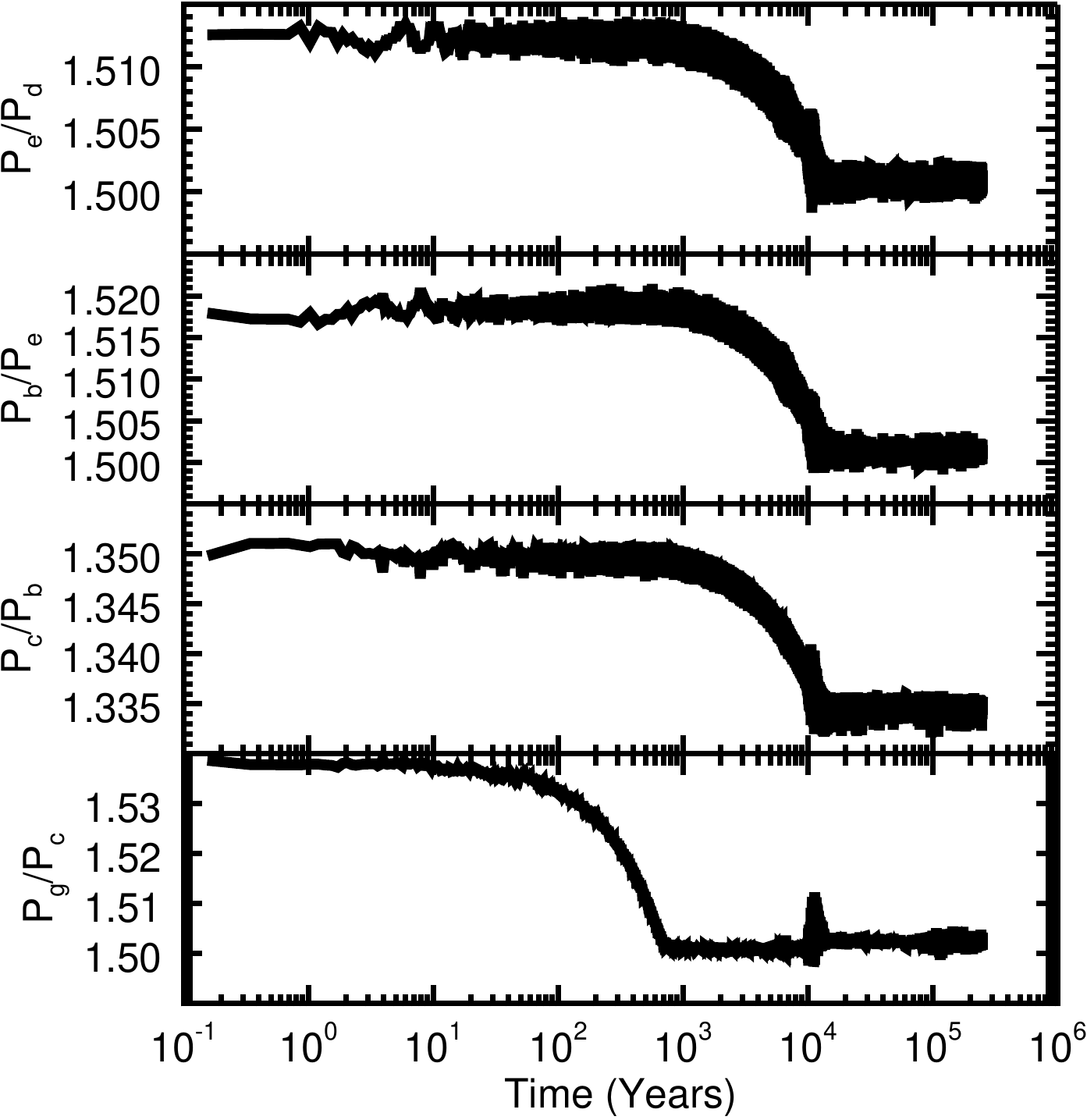}
    \includegraphics[width=0.48\textwidth]{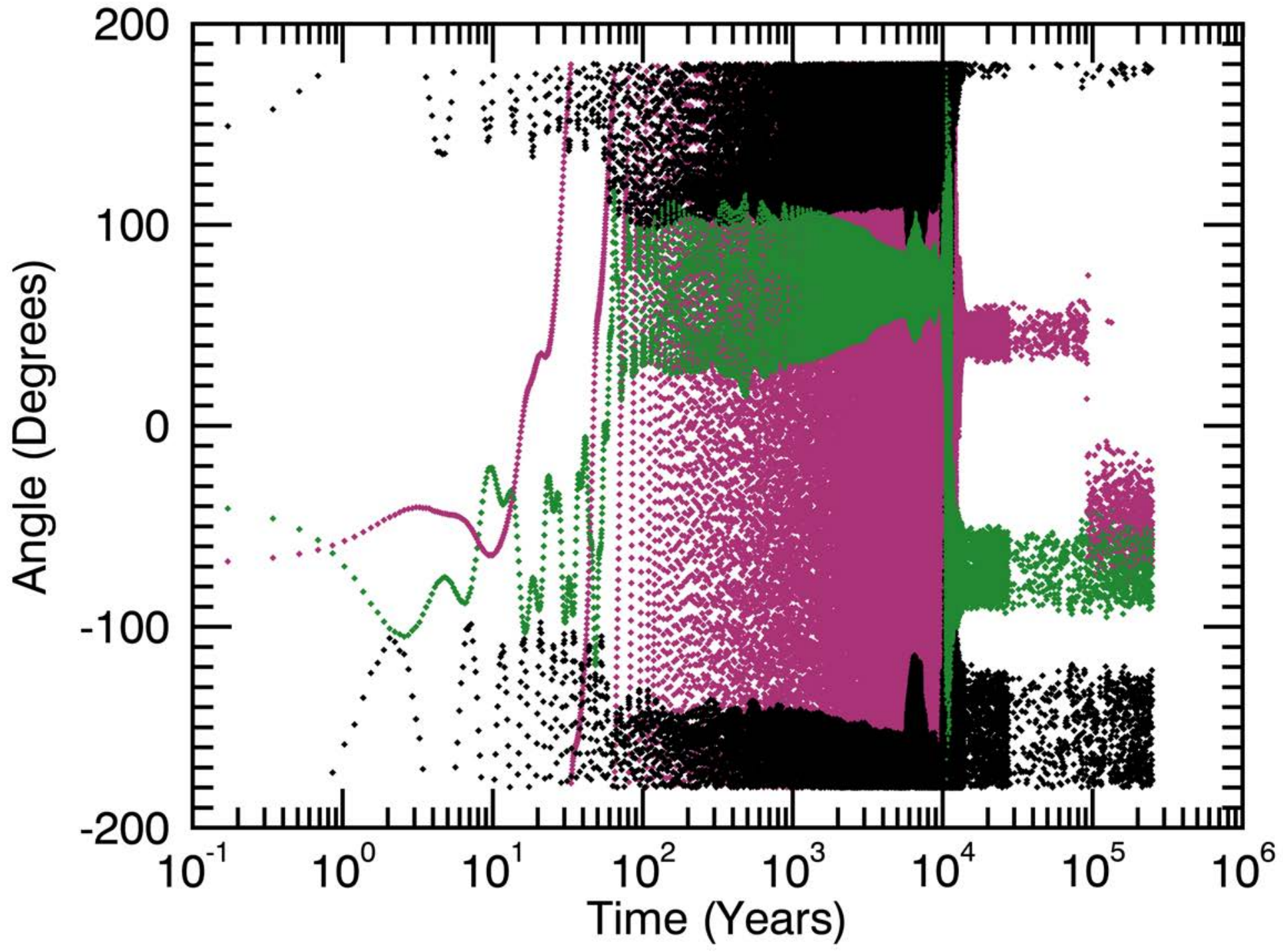}
    \caption{Example of a short scale migration simulation where all 5 planets lock into a resonant chain. Here,
    $\phi_1~=~3\lambda_b - 5\lambda_e + 2 \lambda_d$ (black),
$\phi_2~=~2\lambda_c - 3\lambda_b + 1 \lambda_e$ (green), and
$\phi_3~=~1\lambda_g - 2\lambda_c + 1 \lambda_b$ (purple). Given that  there is a large range of initial conditions that lead to this system being in resonance, we can conclude that this chain is consistent with in situ formation. For this specific example, $\tau_a = 1.5\times10^7$ days and $\tau_e = 3\times10^4$ days, although we explore $10^6~<~\tau_a~<~10^9$ days and $10^2~<~\tau_e~<~10^9$ days.}
    \label{fig:formation}
\end{figure*}

We use the same stellar properties as in our photodynamic model ($M_{\star} = 0.73M_{\odot}$ and $R_{\star} = 0.68R_{\odot}$) and the draw the planetary parameters from our DEMCMC fitting (see Table~\ref{tab:results}). We do not model \pf~as it is not part of the resonant chain. We start all planets at slightly inflated periods ($\sim$10\%) and apply a migration force as semi-major axis and eccentricity damping with timescales of $\tau_a=$ log N$[10^7, 0.7]$ and $\tau_e=$ log N$[10^4.5, 1.0]$ using \texttt{REBOUNDx} \citep{reboundx}. We force the planets to migrate for 5$\times10^6$ days, before turning off migration and integrating the system for another $8.6\times 10^7$ days to verify stability and resonance. We plot the periods, eccentricities, period ratios, and three-body angles of an example simulation where all five outer planets are part of a resonant chain in Figure~\ref{fig:formation}. We also plot the resonant chain outcome as a function of the two damping timescales $\tau_a$ and $\tau_e$ in Figure~\ref{fig:tauK80}, including simulations from \citet{MacDonald2018}.

The full resonant chain of Kepler-80 forms under a large range of initial orbital periods (3-15\%), planetary masses and inclinations, and damping timescales for both semi-major axis ($10^6<\tau_a<10^9$) and eccentricity ($10^2<\tau_e<10^9$)\footnote{These values result from our simulations as well as those from \citet{MacDonald2018}.}. We find, then, that Kepler-80 and its five-planet resonant chain are indeed consistent with in situ formation. We note that we cannot yet conclude that this system \textit{definitely} formed in situ, as the data and model parameters are also consistent with long-scale migration \citep[see discussion in ][]{MacDonald2018}. A further exploration of indicators of in situ vs migration history will be necessary to confirm the formation history of Kepler-80.


\subsection{No correlation between damping timescales}

In exploring the first resonant pair of GJ876, \citet{Lee2002} fix $\tau_a$ and set their eccentricity damping timescales to $\tau_e = \tau_a/K,$ where $K$ is randomly drawn. Many studies since have forced the damping timescales of the semi-major axis ($\tau_a$) and eccentricity ($\tau_e$) to be derived from one another or correlated \citep[e.g., ][]{Tamayo2017}. 

However, given the lack of constraint\footnote{In many isolated cases, $\tau_a$ and $\tau_e$ are correlated, such as in  direct tidal damping \citep[e.g., ][]{Goldreich1966} and tidal torques from a planet embedded in gas \citep{Papaloizou2000}, but these derivations are only for one or two planets and typically assume small eccentricities. In addition, this value of $K$ can be a function of eccentricity, semi-major axis, system age, and/or disk properties.} to $K$ and the growth of computational power in the past few years, these two timescales should be sampled independently from one another. Such an analysis can later lead to interesting discoveries involving how disk and planet properties individually affect planet formation and evolution.

We analyze and expand on the simulations from \citet{MacDonald2018} and look to see if specific damping timescales lead to stability and resonance for Kepler-80. These simulations were performed before the discovery of \pg, and so we only include the four middle planets in our simulations. 

We find that there is no correlation between $\tau_e$ and $\tau_a$, aside from a slight slope on the stability boundary at fast semi-major axis damping. Additionally, we find that Kepler-80 almost always forms a full resonant chain, except in a few cases of fast semi-major axis and eccentricity damping. Some partial resonant chains also form at slow semi-major axis damping, but these simulations often start with at least one planet pair already in resonance.

We also explore similar parameter spaces for resonant system TRAPPIST-1 and for Kepler-60, a three planet system with a suspected resonant chain. We find that the resonant chain of TRAPPIST-1 can form only under high eccentricity damping ($\tau_e<5\times10^4$ days), but can form under a range of semi-major axis damping. We find that most of the simulations, though, result in the system going unstable. 

\begin{figure*}
    \centering
    \includegraphics[width=0.5\textwidth]{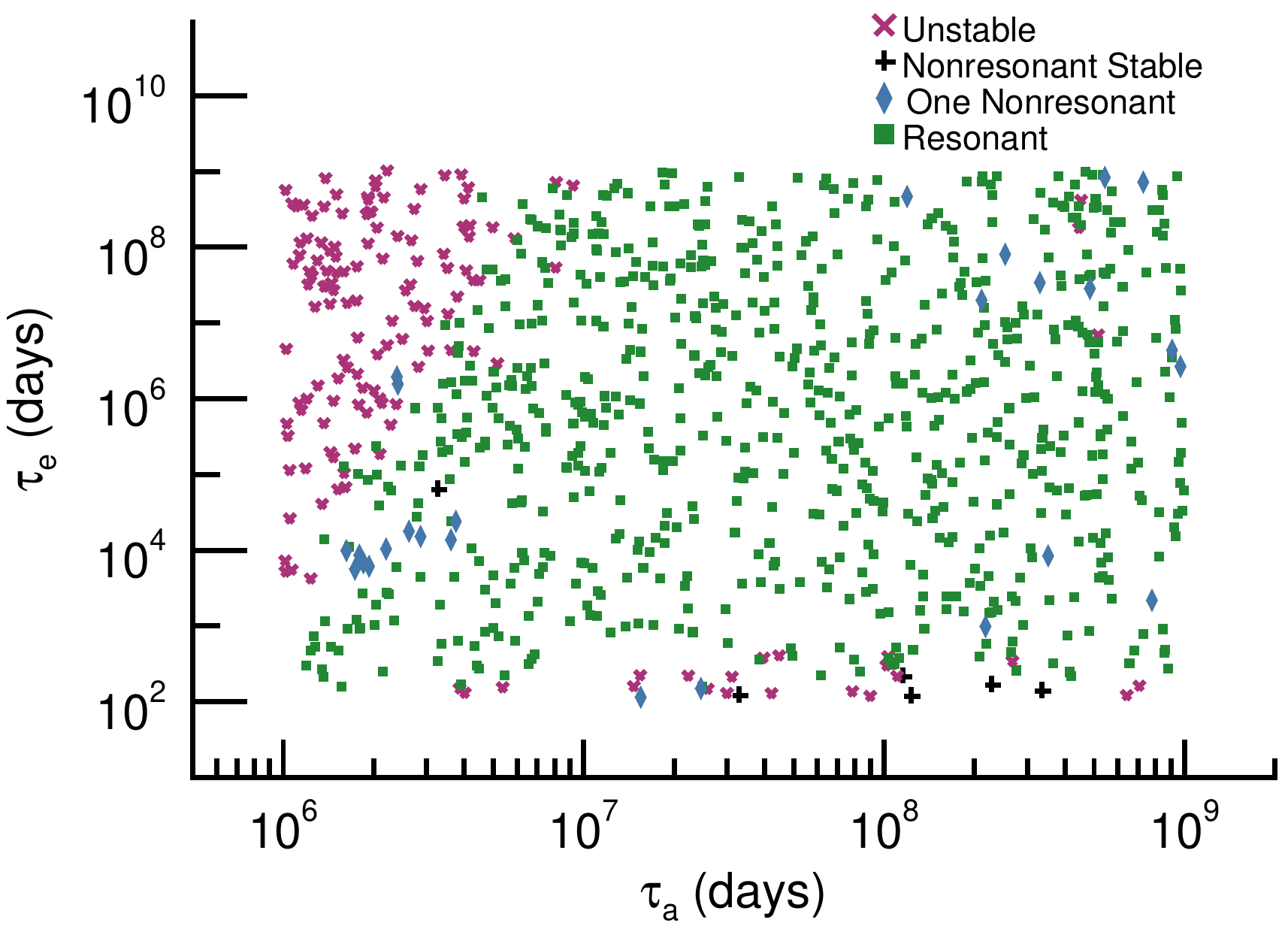}
    \includegraphics[width=0.5\textwidth]{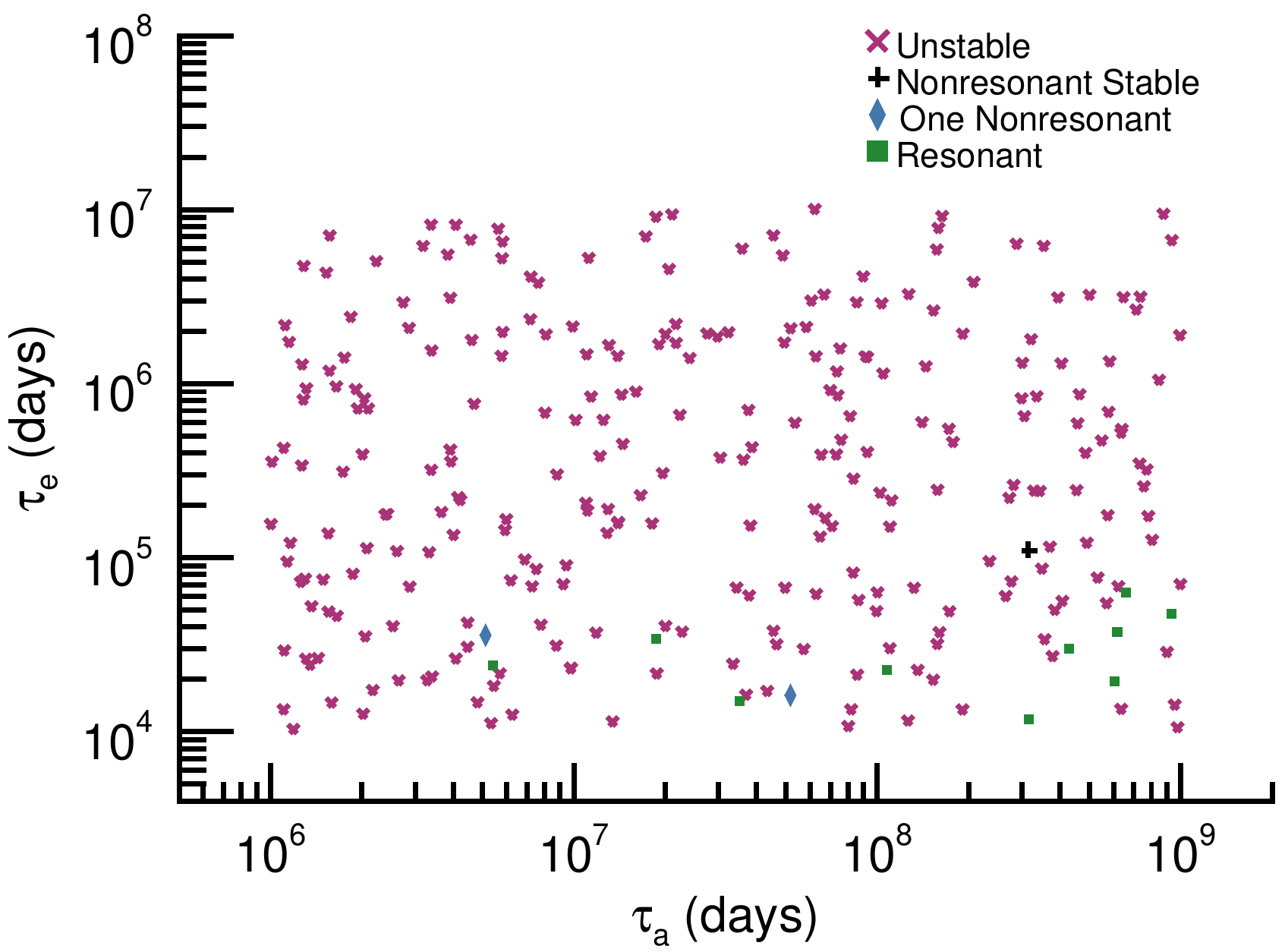}
    \includegraphics[width=0.5\textwidth]{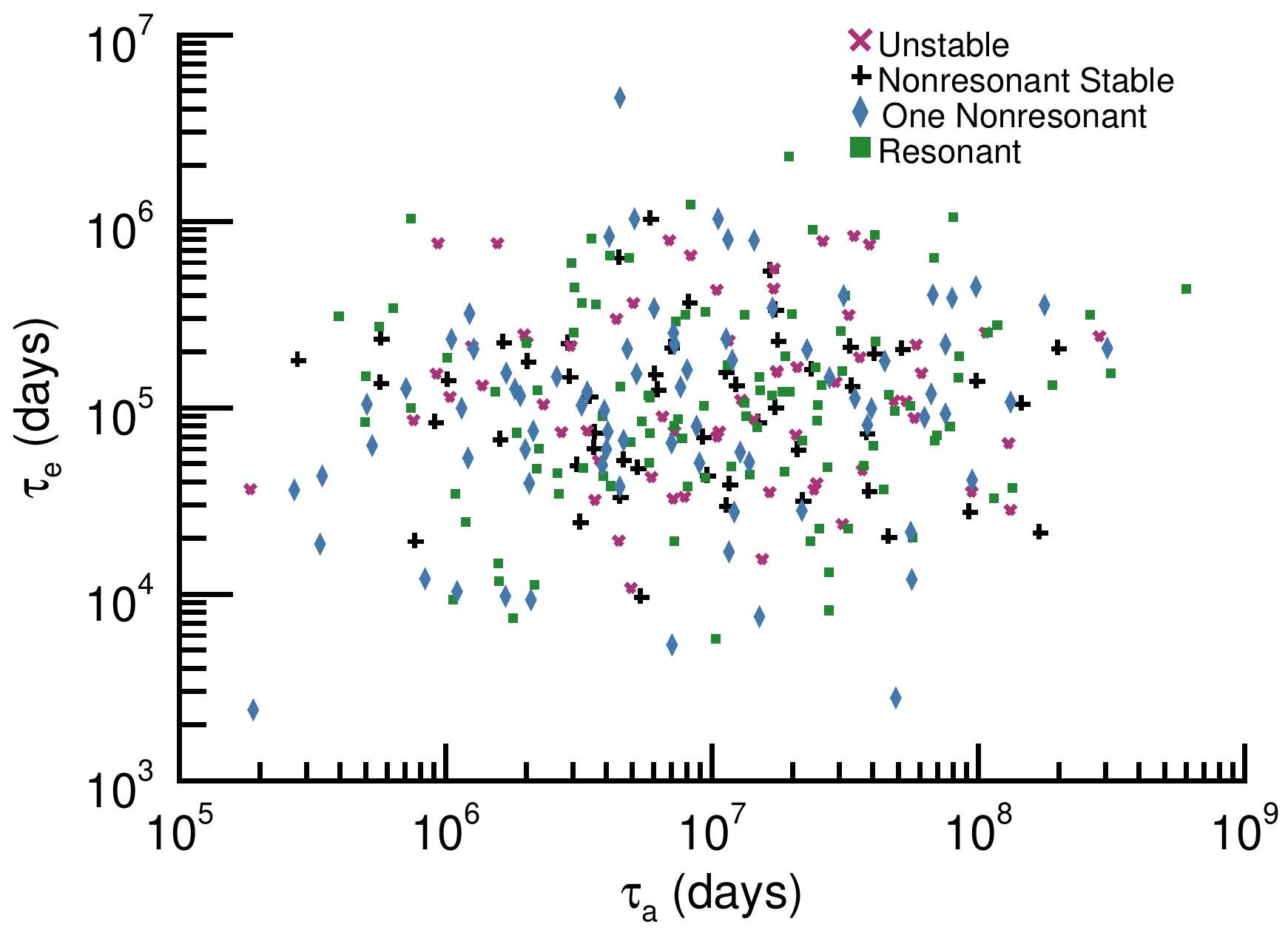}
    \caption{Semi-major axis ($\tau_a$) and eccentricity ($\tau_e$) damping timescales for short scale migration simulations from \citet{MacDonald2018} and from this work, for top) Kepler-60, middle) TRAPPIST-1, and bottom) Kepler-60. We colour the points based on the outcome of the simulation, where purple x's went unstable, black pluses are stable for the full integration time ($9.13 \times 10^7$ days) but have more than one planet out of resonance, blue diamonds have one planet not participating in the resonant chain, and green squares have all planets in the resonant chain. Here, we define a resonant chain as each planet interacting with the others via a librating resonant angle, whether that angle is a two-body, and three-body, or a four-body angle. For all three systems, we find no correlation between $\tau_a$ and $\tau_e$. We find that Kepler-80 typically forms a fully librating resonant chain under a wide range of damping timescales. TRAPPIST-1 is long-term stable if it forms its resonant chain, and this is only possible under high eccentricity damping (small $\tau_e$). We find that Kepler-60 does not need to form a resonant chain to remain long-term stable, even after we force short scale migration. Instead, we find that the system is just as likely to be non-resonant as it is to have only two planets in resonance or all planets in resonance.}
    \label{fig:tauK80}
\end{figure*}

In contrast to TRAPPIST-1, we find that Kepler-60 need not be in a resonant chain to be stable. For a wide range of $\tau_a$ and $\tau_e$, we find that the system is as likely to be non-resonant as it is to have only one resonant pair or to be a fully resonant chain. We plot the results of the short scale migration simulations for all three systems in Figure~\ref{fig:tauK80}. 

Because of the different areas of the $\tau_a$-$\tau_e$ parameter space that can result in stability and resonant chain formation for these three different systems, we caution future studies against forcing a correlation, i.e., $\tau_e = \tau_a/K$. We instead encourage future studies to more fully explore the parameter space for both short-scale and long-scale migration simulations as this can add to the understanding of the initial disk conditions required for the observed system architectures.


\section{Summary and Conclusion}\label{sec:conclusion}

Recently, \citet{Shallue2018} used neural nets to uncover another planet in the system Kepler-80, \pg. This new planet is near the 3:2 period ratio with its neighbor, suggesting that it too is part of the resonant chain confirmed by \citet{MacDonald2016}. Given this new planet, we recharacterize the system. We use PhoDyMM to infer orbital and physical parameters for all six planets simultaneously.  

We find that, although \citet{MacDonald2016} slightly overestimated the masses of the outer two planets, this was likely due to the signal of then unknown planet g, and our resulting masses are consistent with those from \citet{MacDonald2016} within 3$\sigma$. We find that \pc~ and \pb~ require an appreciable atmosphere of $\sim$1-2\% H/He. \pd, \pe, and \pg~ are consistent with a terrestrial composition, although \pd~ requires has a higher Fe/Si fraction than Earth, and \pg~ may require a small, Venus-like atmosphere.

We next integrate forward 1000 of our bestfits and explore the dynamics of the Kepler-80 system. We find that the three- and four-body resonant angles involving \pg~ do not always librate, and the two-body angles for all planet pairs rarely librate. We confirm a four-body resonant chain between planets d, e, b, and c as those associated three-body angles are always librating, but a fully librating five-planet chain only exists in 14\% of our bestfit solutions.

Lastly, we explore a potential pathway for the formation of this system and its resonant chain by performing 100 $N$-body simulations with short scale migration. We find that the chain is consistent with \emph{in situ} formation, although it very well could have formed via other methods. 

\begin{acknowledgments}
MGM acknowledges that this material is based upon work supported by the National Science Foundation Graduate Research Fellowship Program under Grant No. DGE1255832. Any opinions, findings, and conclusions or recommendations expressed in this material are those of the author and do not necessarily reflect the views of the National Science Foundation. DR acknowledges the support of Sean Mills and Daniel C. Fabrycky in developing PhoDyMM. 
\end{acknowledgments}



\bibliographystyle{apj}
\bibliography{bib.bib} 

\end{document}